\documentclass[twocolumn]{aastex631}

\usepackage{comment}

\shortauthors{Rinaldi et al.}

\graphicspath{{./}{figures/}}

\begin{document}

\title{The galaxy starburst/main-sequence bimodality over five decades in stellar mass at $z\approx3-6.5$}

\correspondingauthor{Pierluigi Rinaldi}
\email{rinaldi@astro.rug.nl}

\author[0000-0002-5104-8245]{Pierluigi Rinaldi}
\affiliation{Kapteyn Astronomical Institute, University of Groningen,
P.O. Box 800, 9700AV Groningen,
The Netherlands
}
\author[0000-0001-8183-1460]{Karina I. Caputi}
\affiliation{Kapteyn Astronomical Institute, University of Groningen,
P.O. Box 800, 9700AV Groningen,
The Netherlands
}
\affiliation{Cosmic Dawn Center (DAWN)}

\author[0000-0001-8289-2863]{Sophie E. van Mierlo}
\affiliation{Kapteyn Astronomical Institute, University of Groningen, 
P.O. Box 800, 9700AV Groningen,
The Netherlands
}
\author[0000-0002-3993-0745]{Matthew L. N. Ashby}
\affiliation{Center for Astrophysics $|$ Harvard \& Smithsonian, 
60 Garden St., Cambridge, MA 02138, USA
}
\author[0000-0001-6052-3274]{Gabriel  B.  Caminha}
\affiliation{Max-Planck-Institut f\"{u}r Astrophysik,
Karl-Schwarzschild-Str. 1, D-85748 Garching,
Germany
}
\affiliation{Kapteyn Astronomical Institute, University of Groningen, 
P.O. Box 800, 9700AV Groningen,
The Netherlands
}

\author[0000-0001-8386-3546]{Edoardo Iani}
\affiliation{Kapteyn Astronomical Institute, University of Groningen, 
P.O. Box 800, 9700AV Groningen,
The Netherlands
}
 
\begin{abstract}

We study the relation between stellar mass ($\mathrm{M_{*}}$) and star formation rate (SFR) for star-forming galaxies over approximately five decades in stellar mass ($\mathrm{5.5 \lesssim log_{10}(M_{*}/M_{\odot}) \lesssim 10.5}$) at $z\approx 3 - 6.5$. This unprecedented coverage has been possible thanks to the joint analysis of blank non-lensed fields (COSMOS/SMUVS) and cluster lensing fields (Hubble Frontier Fields) which allow us to reach very low stellar masses.  Previous works have revealed the existence of a clear bimodality in the $\mathrm{SFR - M_{*}}$ plane with a star-formation Main Sequence and a starburst cloud at $z\approx4-5$.  Here we show that this bimodality extends to all star-forming galaxies and is valid in the whole redshift range $z\approx 3 - 6.5$. We find that starbursts constitute at least $\approx 20\%$ of all star-forming galaxies with $\mathrm{M_{*} \gtrsim 10^{9}\;M_{\odot}}$ at these redshifts and reach a peak of 40\% at $z=4-5$. More importantly, 60\% to 90\% of the total SFR budget at these redshifts is contained in starburst galaxies, indicating that the starburst mode of star-formation is dominant at high redshifts. Almost all the low stellar-mass starbursts with $\mathrm{log_{10}(M_{*}/M_{\odot}) \lesssim 8.5}$ have ages comparable to the typical timescales of a starburst event, suggesting that these galaxies are being caught in the process of formation.  Interestingly, galaxy formation models fail to predict the starburst/main-sequence bimodality and starbursts overall, suggesting that the starburst phenomenon may be driven by physical processes occurring at smaller scales than those probed by these models.

\end{abstract}


\section{Introduction} \label{Section_1}
In recent decades, galaxy surveys have constrained several aspects of galaxy evolution up to very high redshifts \cite[e.g., ][]{LeFloch_2005, Ellis_2013, Oesch_2014, Bouwens_2015, Stefanon_2019, Bowler_2020, Bhatawdekar_2021,  Bouwens_2021}.  Much effort has been devoted to constrain galaxy physical properties, such as stellar masses ($\mathrm{M_{*}}$) and star formation rates (SFR). These quantities are fundamental in order to probe the process of gas conversion into stars, i.e., the stellar mass assembly \citep[e.g., ][]{Casey_2012, Huillier_2012, Bauer_2013, Jackson_2020}. The statistical analysis of large galaxy samples revealed a correlation between M$_{*}$ and SFR for star-forming galaxies \citep[e.g., ][]{Brinchmann_2004, Elbaz_2007, Daddi_2007, Noeske_2007}, the so-called {\em galaxy Main Sequence (MS) of star formation}, and the existence of a passive cloud, comprised of galaxies whose instantaneous star-formation rates are negligible with respect to their average past values. These initial works triggered a vast amount of later papers studying galaxy evolution on the  $\mathrm{SFR-M_{*}}$ plane    \citep[e.g., ][]{Peng_2010,  Speagle_2014, Salmon_2015, Santini_2017}.

The existence of a galaxy star-formation MS suggests that similar mechanisms could be responsible for growing low- and high-mass galaxies alike \citep[][]{Noeske_2007}. The MS galaxies grow continuously over a long time period from smooth gas accretion \citep[e.g., ][]{Almeida_2014}. The position of a galaxy on the $\mathrm{SFR-M_{*}}$ plane has been proposed to be strictly correlated with its evolutionary stage \citep[e.g.,][]{Tacchella_2016}, while the intrinsic scatter of the MS suggests some variety in the star-formation histories for galaxies of a given stellar mass \citep[e.g.,][]{Matthee_2019}.

Many studies point out that the normalisation of the $\mathrm{SFR-M_{*}}$ relation increases with cosmic time \citep[e.g.,][]{Whitaker_2012, Whitaker_2014, Iyer_2018}, especially at $z\approx 0-3$, as the gas accretion rate and therefore star formation rate was higher in the past. The relation between M$_{*}$ and SFR is generally parametrised as a power-law of the form log$_{10}$(SFR) = $\alpha$log$_{10}$(M$_{*}$) $+$ $\beta$, where $\alpha$ is the slope and $\beta$ is the intercept. Various studies have been carried out to determine the slope of this relation, and found it to range between 0.6 and 1.0 (see \citet{Speagle_2014} and references therein).

Until recently, only a small fraction of star-forming galaxies were known to be placed significantly above the MS in the $\mathrm{SFR-M_{*}}$ plane, which are the so-called {\em starburst (SB)} galaxies \citep[e.g.,][]{Muxlow_2006, Lee_2017, Orlitova_2020}. The starburst phenomenon is usually driven by a large amount of interstellar gas (mostly in the form of molecular hydrogen), gathered in the galaxy's core. That amount of gas is capable to sustain the typical timescales of the starburst phenomenon ($\mathrm{\approx 10^{7}\;yr}$, \citealp{Heckman_2006}). There is no unique definition of starburst galaxy, rather certain criteria are adopted to identify these sources. For instance, some authors define starbursts as sources which lie N$\sigma$ times above the main sequence (e.g., 4$\sigma$ times in \citealp{Rodighiero_2011}). In this work, we define as a starburst galaxy all those sources with specific SFR (sSFR) $\mathrm{> 10^{-7.60}\; yr^{-1}}$, as proposed by \citet[][]{Caputi_2017, Caputi_2021}. 

Although many theories have been proposed to explain the starburst phenomenon, the nature and growth mechanisms of these galaxies are still under debate. Several works suggest that a violent and large-scale gravitational instability, entirely driven by the self-gravity of the stars, could lead to the starburst phenomenon \citep[e.g., ][]{Inoue_2016, Romeo_2016, Tadaki_2018}. Other works propose that a SB galaxy could consist of many discrete bursts of star formation, probably as a consequence of merging events \citep[e.g.,][]{Lamastra_2013, Calabro_2019}. In contrast with the interpretation that a SB galaxy could be a mere evolutionary stage of MS galaxies displaying a high star formation efficiency, several papers proposed that SB galaxies could be a sort of primeval galaxies with an anomalously high total gas mass \citep[e.g.,][]{Scoville_2014, Genzel_2015, Scoville_2017}. Given their rarity, several works concluded that SB galaxies could have played a minor role in the cosmic star-formation history  \citep[e.g.,][]{Rodighiero_2011, Sargent_2012,  Lamastra_2013}. All these works were solely based on the analysis of relatively massive galaxies with M$_{*}> 10^{10} \, \rm M_\odot$.

A few years ago, \citet{Caputi_2017} discovered the existence of a significant bimodality for star-forming galaxies in the  $\mathrm{SFR-M_{*}}$ plane. This study was based on a sample of prominent H$\alpha$ emitters at $z\approx 4-5$. The bimodality for star-forming galaxies is most evident in the specific star formation rate distribution, which shows two peaks corresponding to the MS and SB cloud. Independently, \citet{Bisigello_2018} analysed a sample of star-forming galaxies at $z=0-3$ and concluded that the fraction of SB becomes increasingly higher with redshift and towards low stellar masses.  These works provide a clear hint that the SB population could have been much more important than previously thought and suggest that investigating low stellar-mass galaxies is essential to fully unveil the SB relevance in the context of galaxy evolution.

Finding the elusive low stellar-mass galaxies can be difficult, though. At low redshifts, faint, low stellar-mass galaxies are observable in deep blank fields. For instance, \citet[][]{Boogaard_2018} made use of the deepest Multi Unit Spectroscopic Explorer (MUSE) observations of Hubble Ultra Deep Field and the Hubble Deep Field South to analyse galaxies reaching stellar masses down to $\mathrm{10^{7}\;M_{\odot}}$ at $z = 0.1 - 0.9$. At high redshifts we need to combine deep fields and gravitational lensing effects offered by massive galaxy clusters, which magnifies the brightness of background sources \citep[e.g., ][]{Pello_2005}. This phenomenon has been successfully exploited to detect intrinsically faint galaxies over a wide redshift range, taking advantage of the flux magnification. As a result, gravitational lensing is a strong tool for better understanding faint, low-mass star-forming galaxies at high redshifts, which would otherwise be inaccessible with current facilities \citep[e.g.,][]{Kikuchihara_2020}. The study of gravitationally lensed, low stellar-mass galaxies at high redshifts indeed suggests that the incidence of starbursts among them could be significantly higher than for higher stellar-mass objects \citep[e.g., ][]{Karman_2017, Caputi_2021}.

Here we present a joint analysis of star-forming galaxies in a cosmological blank field, namely COSMOS/SMUVS \citep[][]{Ashby_2018}, and three lensing cluster fields from the Hubble Frontier Fields (HFF) program \citep[][]{Koekemoer_2016, Lotz_2017}, allowing us to conduct an unprecedented study of the $\mathrm{SFR-M_{*}}$ plane over approximately five decades in stellar mass at $z \approx 3 - 6.5$. This paper is organised as follows. Section \ref{Section_2} provides a brief overview of the datasets used in this work. In Section \ref{Section_3}, we describe the sample selection and the photometry. In Section \ref{Section_4}, we briefly describe the Spectral Energy Distribution (SED) analysis performed on the sample. In Section \ref{Section_5}, we analyse the properties of the lensed Lyman Alpha Emitters (LAEs). In Section \ref{Section_6}, we analyse the $\mathrm{SFR - M_{*}}$ and $\mathrm{sSFR - M_{*}}$ planes, taking into account the HFF and SMUVS galaxies, at  $z \approx 3 - 6.5$. Finally, in Section \ref{Section_7} we present our conclusions. Throughout this paper, we adopt the $\Lambda$-CDM concordance cosmological model (H$_{0}$ = 70 km/s/Mpc, $\Omega_{M}$ = 0.3 and $\Omega_{\Lambda}$ = 0.7). All magnitudes and fluxes are total, with magnitudes referring to the AB system \citep[][]{Oke_1983}. Stellar masses and star formation rates refer to a Chabrier initial mass function (IMF) \citep[][]{Chabrier_2003}.

\section{Datasets} \label{Section_2}
\subsection{The Hubble Frontier Fields}

The HFF program \citep[][]{Koekemoer_2016, Lotz_2017} consists of multi-cycle {\sl Hubble Space Telescope} \citep[{\sl HST,}][]{Fienberg_1986} observations to target six galaxy lensing clusters in parallel with six parallel blank field images ($\approx$ 6 arcmins from the cluster core). The HFF target clusters were selected based on their lensing power strength and their low/moderate zodiacal and Galactic background. The principal scientific aim of the HFF program is to investigate the high-redshift Universe that can only be observed with deep {\sl HST} observations, showing us first clues of the early Universe that, starting next year, we could observe with the {\sl James Webb Space Telescope} \citep[{\sl JWST},][]{Gardner_2006} in more detail.

In this work we consider data from MACS J0416.1-2403 (M0416, R.A. = 04:16:09.89, Dec = -24:03:58.0), Abell 2744 (A2744, R.A. = 0:14:18.78, Dec = -30:23:09.87), and Abell 370 (A370, R.A. = 02:39:52.9, Dec = -01:34:36.5). M0416 is a merging galaxy cluster \cite[][]{Mann_2012} at $z = 0.397$ \citep[][]{Ebeling_2014}. The total mass is $\approx$ 1.2 $\times$ 10$^{15}$ M$_{\odot}$ \citep[][]{Grillo_2015}, if we consider a radius of 950 kpc. A2744 is a massive X-ray luminous merging cluster located at $z = 0.308$ \citep[][]{Abell_1989}, with a virial mass of $\approx$ 1.8 $\times$ 10$^{15}$ M$_{\odot}$ within a radius of 1.3 Mpc. Finally, A370 \citep[][]{Abell_1958} is a galaxy cluster at $z = 0.375$ \cite[][]{Struble_1999}. It is well-known since it hosts the first-ever detected gravitational arc \citep[][]{Soucail_1987b, Soucail_1987}. Therefore, it is one of the best-studied strong-lensing clusters \citep[][]{Medezinski_2010}. Several studies of this system suggest that it has a virial mass $\approx 1 \times 10^{15} \, \rm M_{\odot}$ \citep[e.g.,][]{Umetsu_2011}.

\subsubsection{HST imaging}
{\sl HST} employed 840 orbits (140 orbits for each galaxy cluster) to observe the HFF galaxy clusters, achieving a superb depth of $\approx$ 28.7 $-$ 29 magnitudes ($5\sigma$). Taking into account the gravitational lensing effects, the effective depths of the HFF observations are significantly better ($\approx 30-33$ magnitudes over very small volumes). For all the HFF targets, {\sl HST} obtained images with the Advanced Camera for Surveys (ACS) and Wide Field Camera 3 (WFC3), in a total of seven broad-band filters (F435W, F606W, F814W, F105W, F125W, F140W, and F160W).  We refer the reader to \citet{Koekemoer_2016, Lotz_2017} for further details about the {\sl HST} datasets. These bands provide an amazing opportunity to probe the ultraviolet (UV) part of the rest-frame spectrum of the galaxies analysed in the present work. We obtained all these public data from the \href{https://archive.stsci.edu/pub/hlsp/frontier/}{Mikulski Archive for Space Telescope}. We refer to the v1.0 data release for each galaxy cluster we analyse in this work. The {\sl HST} images have an angular resolution of 0.03 arcsec per pixel and a full width half maximum (FWHM) of 0.20 arcsec in the F160W band, which allows for resolving substructures within some galaxies even at high redshifts.

\subsubsection{VLT/MUSE spectroscopic data}

In this work, we analyse data from MUSE \citep[][]{Bacon_2012}, an instrument mounted on the Yepun telescope (UT4) at the Very Large Telescope (VLT). It allows to observe in two different modes: Wide Field Mode (WFM) and Narrow Field Mode (NFM). In this work, we refer to observations carried out with MUSE WFM. WFM allows for integral field spectroscopy over a field of view of 1 arcmin$^{2}$, providing a spectrum for each $0.2 \times 0.2$ arcsec$^{2}$ pixel element with a Point Spread Function (PSF) of $\approx 0.6-0.8 \, \rm arcsec$. Therefore, MUSE offers an incredible opportunity to blindly (i.e., without pre-selection of targets) look for Lyman-$\alpha$ emitters at $z\approx 2.8 - 6.5$ behind galaxy clusters. It allows to cover a spectral range between 4750 \text{\r{A}} and 9350 \r{A}, with a spectral resolution of $\mathrm{\approx 2.4}$ \text{\r{A}}, reaching LAEs down to $\mathrm{1\times 10^{-18}\;erg\;s^{-1}\;cm^{-2}}$ in 1 arcmin$^{2}$ field with only 4 hours of exposure.

We make use of MUSE observations covering the three galaxy clusters analysed in this work.

M0416 is a well-known galaxy cluster of the HFF program. In this work, we refer to the analysis made by \citet{Vanzella_2020}. It is based on two pointings: the deep pointing as the MUSE Deep
Lensed Field (MDLF) centred in the north-east part of the cluster (17.1h integration time, 0100.A-0763(A), PI: Vanzella) and the observation in the south-west (11-hour integration, 094.A-0525(A), PI: Bauer). We refer the reader to \citet{Bergamini_2020} for detailed information about the lensing model adopted to analyse this cluster.

A2744 was observed with MUSE between September 2014 and October 2015 as part of the Guaranteed Time Observations (GTO) Program (094.A-0115, PI: Richard). The MUSE pointings cover a  $2 \times 2$ arcmin$^{2}$ mosaic with the purpose of covering the entire multiple-image area. The entire area was split into four quadrants, which have been observed for a total of  3.5, 4, 4 and 5 hours. Additional time (2 hours) was used to observe the centre of the galaxy cluster. The entire MUSE mosaic overlaps all 7 HFF bands we adopt in this work. For detailed information about data reduction and the adopted lensing model, we refer the reader to \citet{Mahler_2018}.

A370 was observed with MUSE using a large mosaic covering $\approx$ 4 arcmin$^{2}$. This mosaic (096.A-0710(A), PI: Bauer) is an extension of an initial GTO program (094.A-0115(A), PI: Richard), since it is focused on the central part of the galaxy cluster. The entire area covered by the mosaic is $2 \times 2$ arcmin$^{2}$, centred on the core of the cluster, which allows to cover nearly the entire multiple-image area. The mosaic comprises 18 hours of on-source exposure, which have been taken from November 2014 to September 2016. For information about data reduction and the adopted lensing model, we refer the reader to \citet{Richard_2021} which provides an updated version of the redshift catalogue obtained by \citet{Lagattuta_2019}.

\subsection{The COSMOS/SMUVS survey}

As a complement, we also consider deep imaging data from the COSMOS field \citep[][]{Scoville_2007}. By design, these data allow us to explore a very different region in parameter space, as they cover a much wider area than the HFF, but are about three magnitudes shallower. Therefore, these blank fields are useful to probe the high-mass end, not probed by the lensed fields.

The Spitzer Matching survey of the Ultra-VISTA ultra-deep stripes \citep[SMUVS,][]{Ashby_2018} is an Exploration Science Program which collected infrared imaging with {\sl Spitzer}'s data \citep[][]{Werner_2004} with the Infrared Array Camera \citep[IRAC, ][]{Fazio_2004} at 3.6 and 4.5~$\rm \mu m$ over $0.66 \, \rm deg^2$ of the COSMOS field. The region covered by SMUVS corresponds to the part of the COSMOS field with deepest near-IR data from the UltraVISTA program \citep{McCracken_2012} and optical Subaru data \citep{Taniguchi_2007}. The SMUVS data has an average integration time of $\approx$ 25 h/pointing and  reaches 80\% completeness at $\approx 25.5$~mag, both in the 3.6 and 4.5~$\rm \mu m$ filters \citep{Deshmukh_2018}.

In this work we make use of the SMUVS galaxy catalogue obtained by \citet{Deshmukh_2018} and updated by \citet{van_Mierlo_2022}. The SMUVS galaxy catalogue contains a total of $\approx 300,000$ {\sl Spitzer} sources extracted using UltraVISTA $HK$-band-selected galaxies as priors, and includes 28-band photometry from the $U$ band through 4.5~$\rm \mu m$. Briefly, \citet[][]{Deshmukh_2018, van_Mierlo_2022} adopted the code \texttt{LePHARE} \citep[][]{LePhare_2011} to perform the SED fitting. They made use of a series of synthetic templates (with solar and sub-solar metallicities) from the \citet[][]{BC_2003} library, adopting a simple stellar population and different exponentially declining star formation histories with star formation timescales $\tau$ = 0.01, 0.1, 0.3, 1.0, 3.0, 5.0, 10.0, and 15 Gyr. Each synthetic spectrum is attenuated using the reddening rule proposed by \citet[][]{Calzetti_2000}, leaving the color excess as a free parameter with values E(B-V)=0.0–1.0 in steps of 0.1. We refer the reader to \citet[][see Section 3.1]{Deshmukh_2018} for more details about how this catalogue has been obtained. The version of the SMUVS catalogue that we consider here has been updated using the latest UltraVISTA data release (DR4).

In this work we only consider star-forming galaxies in SMUVS between $z = 2.8$ and $z = 6.5$, i.e., passive galaxies as well as active galactic nuclei (AGNs) are excluded. To remove passive galaxies, which account for $\approx 5\%$ of SMUVS sources between $z = 2.8$ and $z = 6.5$, a combination of SED fitting analysis and colour criteria has been adopted. We refer the reader to \citet[][]{Deshmukh_2018} for more details. In order to exclude any possible AGN contamination, we cross-match our SMUVS sources with the {\sl Spitzer} MIPS 24 $\mu$m catalogue \citep[][]{Sanders_2006}, adopting a radius of 2 arcsec. We find that $\approx 1\%$ of the SMUVS sources have a flux $S_\nu(24 \mu \rm m) > 0.2 \, \rm mJy$, which at high redshifts are likely AGNs \citep{Stern_2005}. As a double check, we cross-matched those sources with the X-ray catalogue from  \citet[][]{Civano_2016}. We find that most of the sources with $S_\nu(24 \mu \rm m) > 0.2 \, \rm mJy$ are X-ray detected. Therefore, we decide to precautionary remove all of them. Finally, we exclude $\approx 2\%$ of the total SMUVS galaxies which show significant X-ray detections in \citet[][]{Civano_2016} and are therefore likely AGNs.

\section{Sample selection and photometric analysis}\label{Section_3}
As our main goal here is to study star formation over five decades in stellar mass at $z \gtrsim 3$, we consider two complementary galaxy samples: one consisting of MUSE spectroscopically-confirmed Ly$\alpha$ emitters in the three lensing clusters, which mostly span low stellar masses ($\mathrm{log_{10}(M_{*}/M_\odot) \approx 6 - 8.5}$) at $z=2.8-6.5$; and the sample of {\em all} star-forming SMUVS galaxies at the same redshifts, which mostly have $\mathrm{log_{10}(M_{*}/M_\odot) > 8.5}$. The properties of the latter have been presented and discussed in \citet{Deshmukh_2018}. In Sec. 3.3 we summarise the main details of Deshmuhk et al's sample selection and analysis, but for the lensed Ly$\alpha$ emitters, we present a more complete description of the photometric measurements and property derivation, which we have obtained independently of other authors who analysed partly overlapping galaxy samples \citep[e.g.,][]{Merlin_2016, Santini_2017}. 

\subsection{Selection of Ly$\alpha$ emitters at $z>2.8$ in the HFF}
\begin{figure}[t]
    \centering
    \includegraphics[width = 0.49 \textwidth, height = 0.30 \textheight]{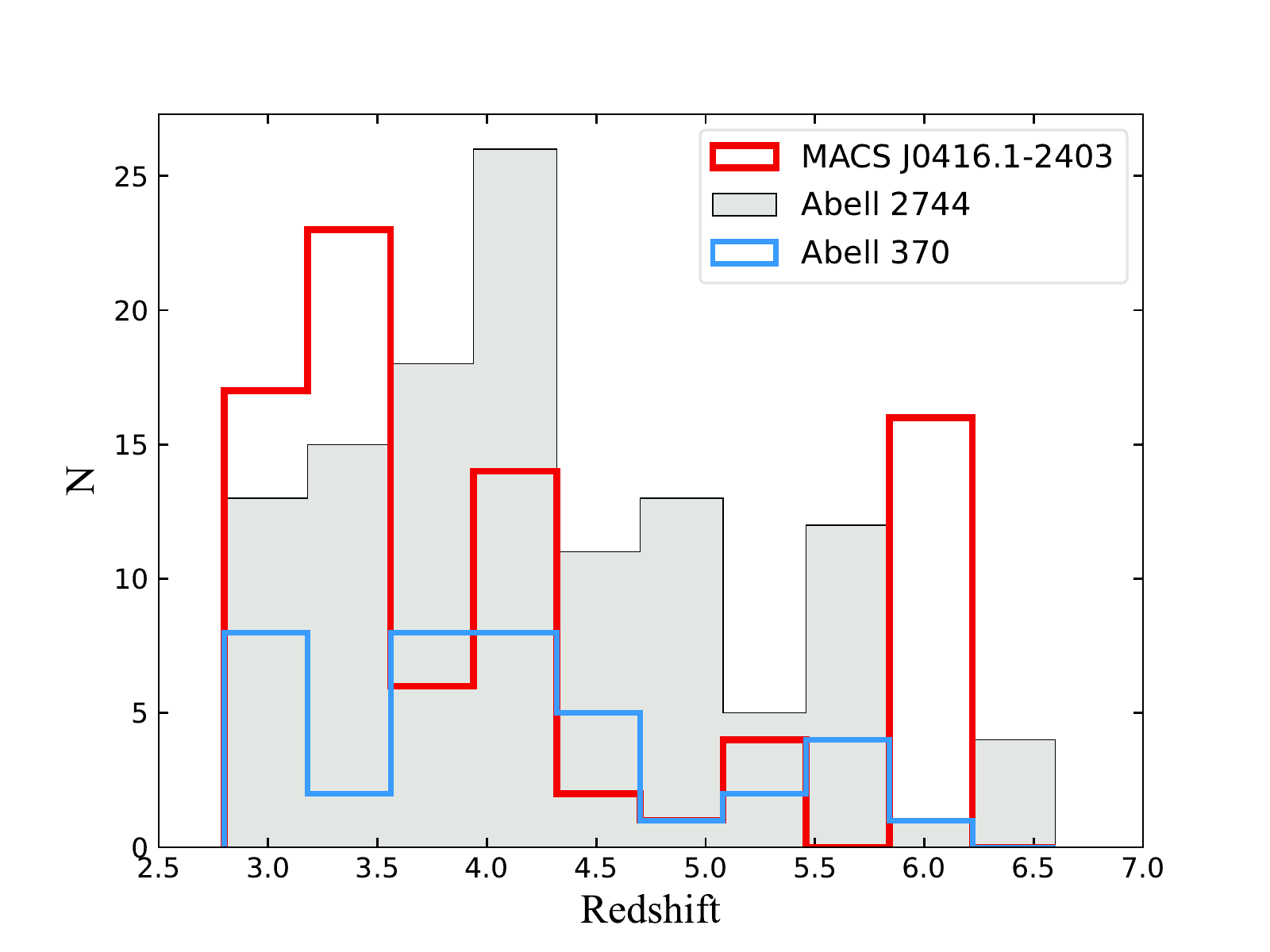}
    \caption{The spectroscopic redshift distribution of analysed galaxies in each HFF cluster we studied here. Galaxies with multiple lensing images have been considered only once.}
    \label{Fig_1}
\end{figure}

In each of the three HFF lensing clusters that we consider here we have selected Lyman Alpha Emitters at $z\gtrsim2.8$ using the available MUSE data and keeping only sources with a robust spectroscopic redshift determination (Table \ref{Table_1}).  Our final sample (M0416 + A2744 + A370) contains 356 imaged sources in total. These correspond to 240 different lensed galaxies. Among these 240 background sources there are 176 with \textit{single} images and 64 objects which have multiple images. The redshift distribution of the analysed sample is shown in Fig. \ref{Fig_1}.

\begin{deluxetable}{cchlDlc}
\tablenum{1}
\tablecaption{Number of Ly$\alpha$ emitters at redshift $2.8 - 6.5$ in the HFF galaxy clusters\label{Table_1}.}
\tablewidth{0pt}
\tablehead{
\colhead{Galaxy cluster} & \colhead{Number of sources}
}
\startdata
MACS J0416.1-2403 & 134 \\
Abell 2744 & 139 \\
Abell 370 & 83 \\
\enddata
\tablecomments{The number of LAEs ($z = 2.8 - 6.5$)  we studied in each galaxy cluster. All these sources have a good quality flag (QF $>1$), i.e., sources have either QF = 2 or QF = 3. QF = 2 refers to a probable redshift with a precision less than $\delta z = 0.001$ given by features less strong but still clearly identifiable. QF = 3 refers to a secure redshift with multiple prominent spectral features or one strong feature (such as Ly$\alpha$). The total amount of all these sources is 356. These correspond to 240 different lensed galaxies. Among them, there are 176 galaxies with a single image and 64 objects with multiple images.}
\end{deluxetable}

\subsection{HST photometry of the HFF  Ly$\alpha$ emitters}
We used the software Source Extractor \citep[\texttt{SExtractor},][]{SExtractor} to detect the MUSE-selected Ly$\alpha$ emitters and measure their photometry in the seven available {\sl HST} broad-bands. We ran \texttt{SExtractor}  in dual-image mode, using an ultra-deep detection image (combining data in multiple filters) as the detection image, and measured fluxes on each band separately. In order to maximise the number of detected Ly$\alpha$ emitters, we chose a {\em hot}-mode configuration in \texttt{SExtractor}, following the method proposed by \citet{Galametz_2013}, which is  optimised for detecting faint sources with small sizes.

\begin{figure*}[ht!]
    \centering
    \includegraphics[width=\textwidth]{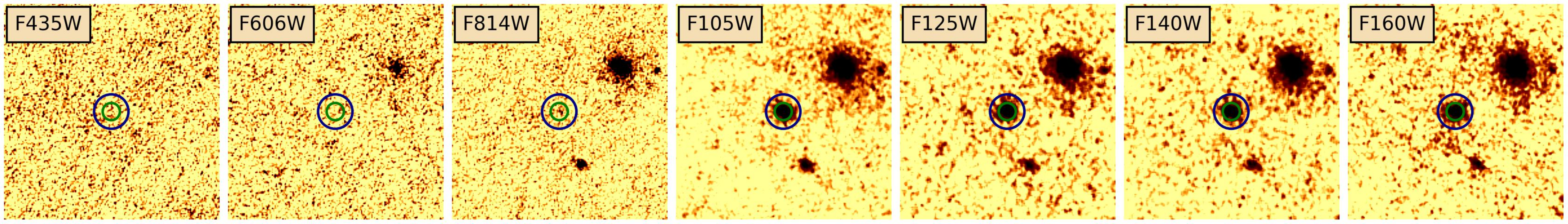}
    \includegraphics[width=\textwidth]{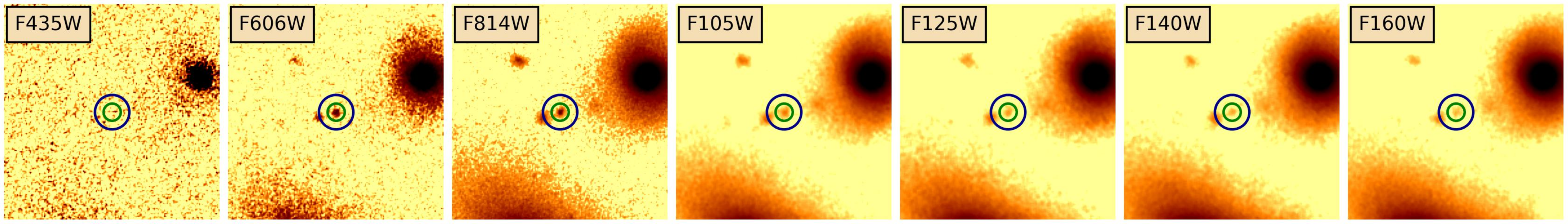}
    \caption{{\sl HST} postage stamps ($\mathrm{5\times5\;arcsec^{2}}$) of two Ly$\alpha$ emitters in A2744. In both cases, we show all the filters considered in this work. In each panel the blue circle indicates an aperture size of 0.8 arcsec diameter, while the green one refers to an aperture size of 0.4 arcsec diameter. For the source shown in the top panels an aperture of 0.8 arcsec is necessary to measure all the galaxy light. On the contrary, for the case at the bottom, an aperture diameter of 0.4 arcsec is sufficient and prevents light contamination from nearby objects.} 
    \label{Fig_2N}
\end{figure*}

We measured each source photometry adopting circular apertures of 0.4 arcsec diameter in \texttt{SExtractor}. This small aperture size is necessary to avoid contamination from close neighbours (in projection) and from intra-cluster light (ICL). As most of our targets are very compact, this aperture size appears to be optimal. However, after a careful visual inspection, we realised that there are some sources for which an aperture size as large as 0.8 arcsec diameter is necessary in order to encompass all the galaxy light (Fig. \ref{Fig_2N}). In all cases, we corrected our aperture fluxes to total using the curve of growth of non-saturated stars in the field.
A minority of sources appear significantly stretched by the lensing effects and, for these, circular aperture photometry is clearly not suitable. For these sources, instead, we measured Kron aperture photometry \citep[i.e., \texttt{MAG\_AUTO} in \texttt{SExtractor},][]{Kron_1980}. In non-crowded regions of the {\sl HST} images, we also compared the \texttt{MAG\_APER} and \texttt{MAG\_AUTO} of each of our targets. If \texttt{FLUX\_AUTO} is greater than \texttt{FLUX\_APER} corrected with the aperture correction factor, then we kept \texttt{FLUX\_AUTO}. Otherwise, we kept \texttt{FLUX\_APER}. In this way, we constructed an optimised version of the photometric catalogue for each galaxy cluster.

Finally, all our fluxes have been corrected for Galactic extinction\footnote{We used the tool available at \url{https://irsa.ipac.caltech.edu/applications/DUST/}.}.   We recover Galactic extinction values which are in a perfect agreement with those present in Table 5 of \citet{Shipley_2018}.

Since \texttt{SExtractor} generally underestimates photometric errors \citep[e.g.,][]{Sonnett_2013}, we decided to manually set a minimum error of 0.05 magnitudes for all those detections with a photometric error less than that value. Indeed, it represents the minimum systematic error for {\sl HST} imaging data.

We performed a positional cross-match between our photometric catalogues and the initial MUSE samples. To do that, we adopted a maximum allowed separation of 0.5 arcsec. In that way, we are able to identify any possible Ly$\alpha$-emitter counterpart. The percentages of recovered sources are listed in Table \ref{Table_2}. We carefully inspected all the sources that did not yield any {\sl HST} match to understand why they are not detected by \texttt{SExtractor}. We noticed that, in most cases, they are too faint to be observed even in the ultra-deep {\sl HST} imaging. In other cases, they are close in projection to bright objects (e.g., a bright cluster member) which prevents the individual detection of the faint sources.

In order to perform an independent check of our \texttt{SExtractor} photometry, we compared our photometry with that from \href{http://www.astrodeep.eu/frontier-fields/}{ASTRODEEP} \citep[][]{Merlin_2016}. In particular, we tested our procedure for M0416 and A2744 because they are the only two galaxy clusters that we have in common with them. To do that, we cross-matched our sources with their catalogues, considering an allowed maximum separation of 0.5 arcsec. Despite the different techniques used by the ASTRODEEP group to perform their photometry, our {\sl HST} photometry is in agreement within the error bars for the vast majority of matched galaxies.  We inspected the few galaxies ($\approx 4\%$ in each galaxy cluster we compared) that yielded significant differences in the photometric measurements: in all cases these sources are close in projection to the cluster centre. The differences in the measured photometry are therefore not surprising, given that the ASTRODEEP measurements include previous cleaning for ICL.  On the one hand, these sources are only $\approx 3\%$ (in the F105W band\footnote{We refer to the F105W band because most of the sources are detectable in that filter.}) of the total sample (M0416 + A2744 + A370). On the other hand, the photometric differences we measured may be due to the different approach used by the ASTRODEEP group. Despite the difference with their photometric measurements, we opted to keep these sources in our study since we do not really know how the ICL cleaning technique could affect the photometry of those sources.

Furthermore, we performed a visual inspection of each source we found in each galaxy cluster to establish the goodness of the cross-matching we carried out with the initial MUSE samples. In that way, we noticed that, in some cases, the centroid of the Ly$\alpha$ emission line falls between two or more sources. Given the size of the MUSE PSF ($\approx 0.6-0.8 \, \rm arcsec$), we are not able to disentangle which of the {\sl HST} sources is the right counterpart. In these cases, we considered all the objects within a radius of 0.4 arcsec as counterparts to the Ly$\alpha$ emitter. For those sources, we assumed both the same redshift and magnification factor ($\mu$) as for the Ly$\alpha$ emitter. Specifically, we found five objects with multiple ($\geq 2$) {\sl HST} counterparts in M0416, six cases in A2744, and only one case in A370.

\begin{deluxetable}{cchlDlc}[ht!]
\tablenum{2}
\tablecaption{The percentages of the sources successfully cross-matched with the initial MUSE samples\label{Table_2}.}
\tablewidth{0pt}
\tablehead{
\colhead{Galaxy cluster} & \colhead{Recovered sources}
}
\startdata
MACS J0416.1-2403 & 88\% \\
Abell 2744 & 81\% \\
Abell 370 & 84\% \\
\enddata
\tablecomments{These percentages refer to the total amount of sources we found in each galaxy cluster, i.e., both single and multiple images.}
\end{deluxetable}

\subsection{Photometry of the SMUVS galaxies}
The construction of the SMUVS galaxy photometric catalogue, which contains fluxes in a total of 28 filters from $U$ through $4.5 \, \rm \mu m$,  has been thoroughly explained in \citet{Deshmukh_2018}. Here we only summarise the main information and refer the reader to that paper for further details.

To measure photometry in every band, except the {\sl Spitzer} bands, \texttt{SExtractor} has been run in dual-image mode, using the UltraVISTA $HK_s$ stack mosaic as detection image. Here we consider an updated version of the SMUVS catalogue, obtained using the latest UltraVISTA release (DR4) as a starting point for the catalogue construction \citep[][]{van_Mierlo_2022}. The {\sl Spitzer} IRAC photometry has been obtained using a PSF-fitting technique, using the position of the $HK_s$ sources as priors. This technique is strictly valid only for point-like sources (see Fig. 25 of \citet[][]{Ashby_2013} for detailed information about how point-like IRAC-detected sources are), but in the {\sl Spitzer} images this is indeed the case for the vast majority of galaxies at $z>2$, given the IRAC PSF size (FWHM $\approx 1.9$ arcsec) \citep[e.g.,][]{McCracken_2012, Ashby_2015, Laigle_2016}.

All SMUVS photometric measurements have been obtained in circular apertures (2 arcsec diameter), corrected to total, and corrected for Galactic extinction.

\section{SED fitting analysis of the HFF Ly$\alpha$ emitters}\label{Section_4}
As the SED fitting and redshift derivation of the SMUVS sources has been thoroughly explained by \citet{Deshmukh_2018}, here we only explain in detail our SED fitting analysis for the HFF Ly$\alpha$ emitters.

Note that, while the SED fitting of the SMUVS galaxies is based on 28 filters, we only use 7 {\sl HST} bands from the HFF program to perform the SED fitting of the Ly$\alpha$ emitters. In spite of this difference, the SED fitting quality is still very good in most cases, as for the Ly$\alpha$ emitters the redshifts are securely known in advance from the spectroscopic determinations (Fig. \ref{Fig_1}).

We performed the SED fitting and derived the properties of our Ly$\alpha$ emitters using the code \texttt{LePHARE}. As a starting point, we adopted the same setup as for SMUVS \citep[][]{Deshmukh_2018} to configure \texttt{LePHARE}. We made use of a galaxy template library with the following set of star formation histories (SFHs):
\begin{itemize}
    \let\labelitemi\labelitemii
    \item A standard exponentially declining, as known as "$\tau$-model", in which the star formation rate (SFR) is SFR($t$) $\propto$ exp$^{-(t-t_{0})/\tau}$. In particular, we adopted the following $e$-folding timescales ($\tau$) in Gyr: 0.01, 0.1, 0.3, 1, 3, 5, 10, 15;
    \item An instantaneous burst, adopting a simple stellar population (SSP) model, which means that a single instantaneous burst of star formation
    took place at time $t$: SFR($t$) $\propto \delta(t)$. 
\end{itemize}
We adopted the stellar population synthesis (SPS) models from \citet{BC_2003} (hereafter BC03) based on a Chabrier IMF \citep[][]{Chabrier_2003}, considering two different values for the metallicity: solar metallicity (Z$_{\odot}$ = 0.02) and a fifth of solar metallicity (Z = 0.2Z$_{\odot}$ = 0.004).  Since we have high-redshift galaxies (at $z \approx 2.8-6.5$), we chose to expand the range of ages used in \citet{Deshmukh_2018} in order to include younger ages as low as 1 Myr. Including these younger ages in our SED fitting analysis prevents an accumulation of results at the youngest allowed age. In order to take into account the effects of internal dust extinction, we convolved the model templates with the \citet[][]{Calzetti_2000} reddening law, with the extrapolation proposed by \citet{Leitherer_2002} at shorter wavelengths. We considered colour excess values between 0 $\leq$ E(B$-$V) $\leq$ 1.0, with a step of 0.1.

In addition, we computed upper limits in every band where \texttt{SExtractor} did not detect any source. We obtained the r.m.s. ($1\sigma$) of the local background by doing statistics on the flux of 50  empty apertures randomly placed in the background around each source. For \texttt{LePHARE}, we considered a 3$\sigma$ upper limit for the flux in the corresponding filter and chose the option in which \texttt{LePHARE} ignores any template that produces a flux above that limit.   Nevertheless,  in a minority of cases  ($\approx 3\%$)  the ICL contaminates the light of sources especially in the reddest filters. In these cases, we ignored the photometry in those filters in the SED fitting (we used -99 in \texttt{LePHARE}, which indicates a lack of photometric information).

\section{Properties of the HFF Ly$\alpha$ emitters}\label{Section_5}
\subsection{SED-derived properties}
\begin{figure}[t!]
    \centering
    \includegraphics[width = 0.49 \textwidth, height = 0.30 \textheight]{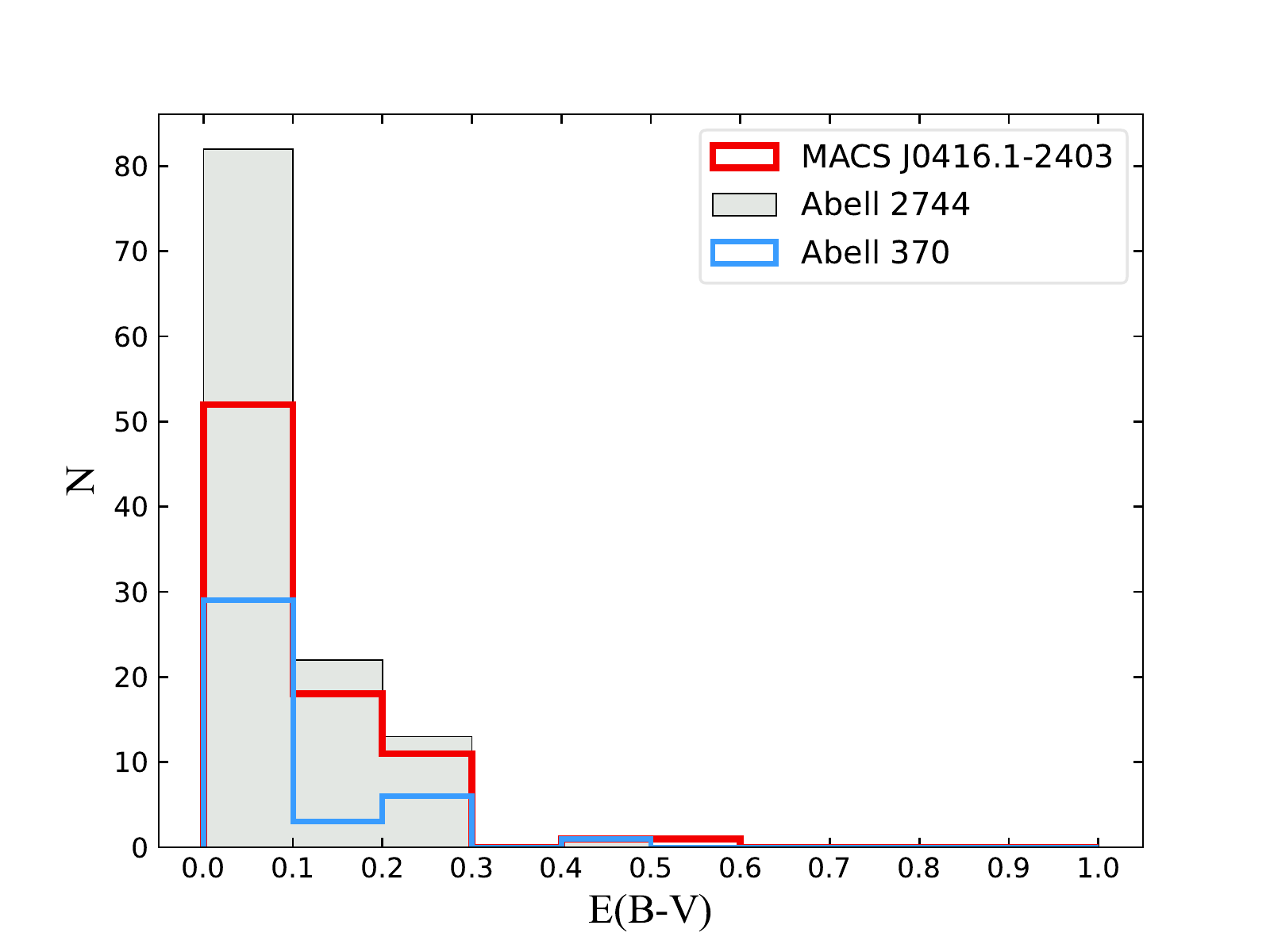}
    \caption{Distribution of colour excess obtained through the SED fitting.}
    \label{Fig_3}
\end{figure}

\texttt{LePHARE} returns the best-fit SED and derived parameters for every Ly$\alpha$ emitter. As stated above, our total sample contains 356 sources, corresponding to 240 different galaxies. In all the following analysis, we will consider only one set of best-fit properties per galaxy, which for the sources with multiple images means that we choose only the best of the best-fit results  (i.e., that with the lowest reduced $\chi^2$).\\
We find that $\approx$ 56\% of our sample has a sub-solar metallicity (0.2 Z$_{\odot}$) and $\approx$ 44\% has a solar one (Z$_{\odot}$). We do not find any correlation between the best-fit metallicity and other parameters considered in our analysis. We find that the best-fit colour excess values range from 0.0 to 0.5. In particular, $\approx$ 68\% of our  sources have E(B-V) = 0.0 (Fig. \ref{Fig_3}). We find that our values of E(B-V) are in agreement with what has been measured in other populations of LAEs \citep[e.g., ][]{Karman_2017, Rosani_2020}.

\begin{figure}[t!]
    \centering
    \includegraphics[width = 0.49 \textwidth, height = 0.30 \textheight]{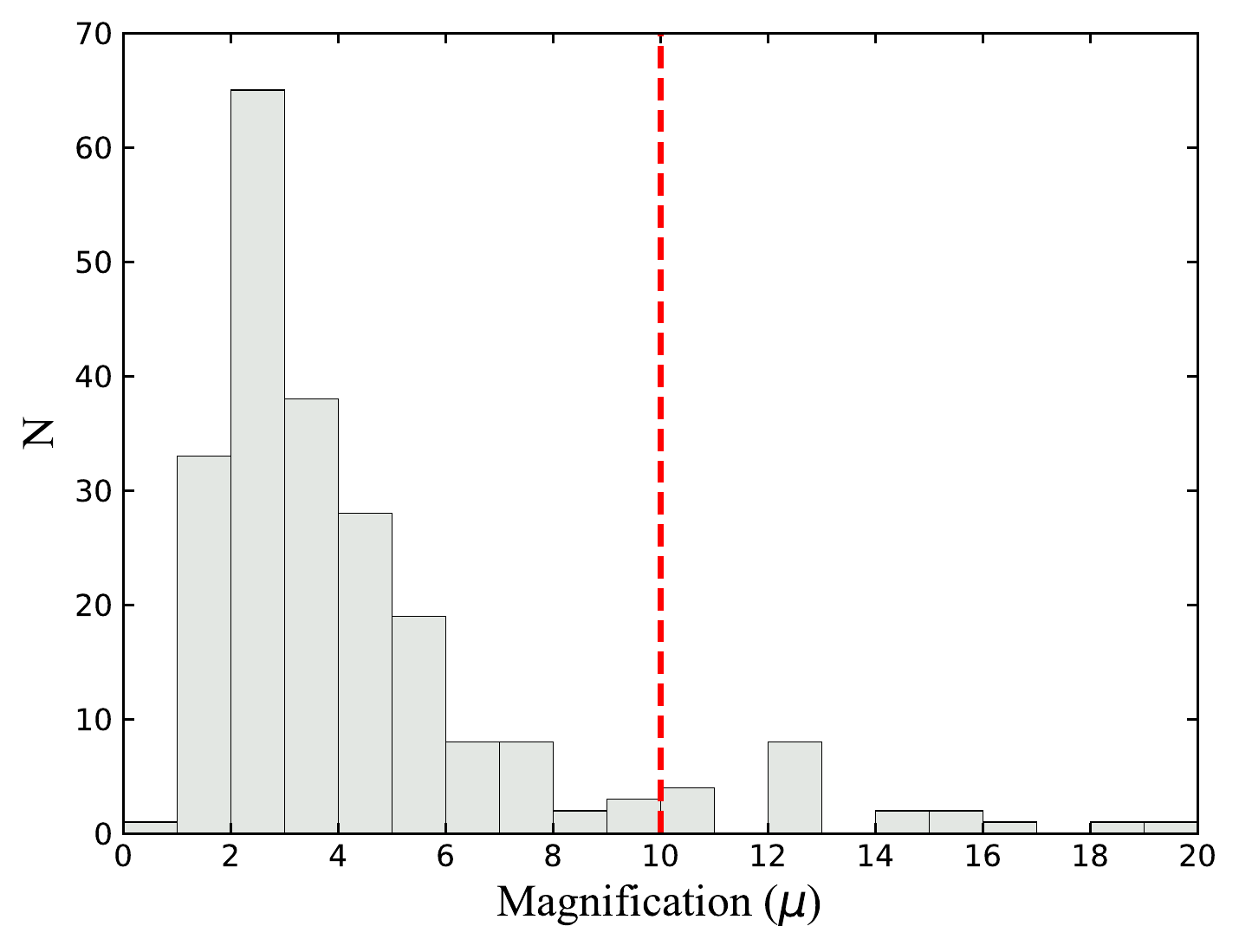}
    \caption{The distribution of the magnification factors for all the HFF galaxy clusters we studied in this work. The red vertical line refers to a limit above which we inspected all the galaxies with $\mu >$ 10 in order to analyse if their presence could impact on the SFR $-$ M$_{*}$ plane or not. To better visualise the distribution, we only show $\mu$ up to 20. There are 33 sources with $\mu$ $>$ 10 ($\approx 14\%$ of the analysed sample).}
    \label{Fig_4}
\end{figure}

\texttt{LePHARE} also determines the best-fit stellar mass for each galaxy. We correct these values for lensing magnification using the lens model of \citet{Bergamini_2020} for M0416, \citet{Mahler_2018} for A2744, and \citet{Richard_2021} for A370.
The magnification factors cover a wide range, with the highest value $>$ 200. The bulk of the sources have a magnification factor $<$ 20. The distribution of magnification values is shown in Fig. \ref{Fig_4}.

\begin{figure}[t!]
    \centering
    \includegraphics[width = 0.49 \textwidth, height = 0.30 \textheight]{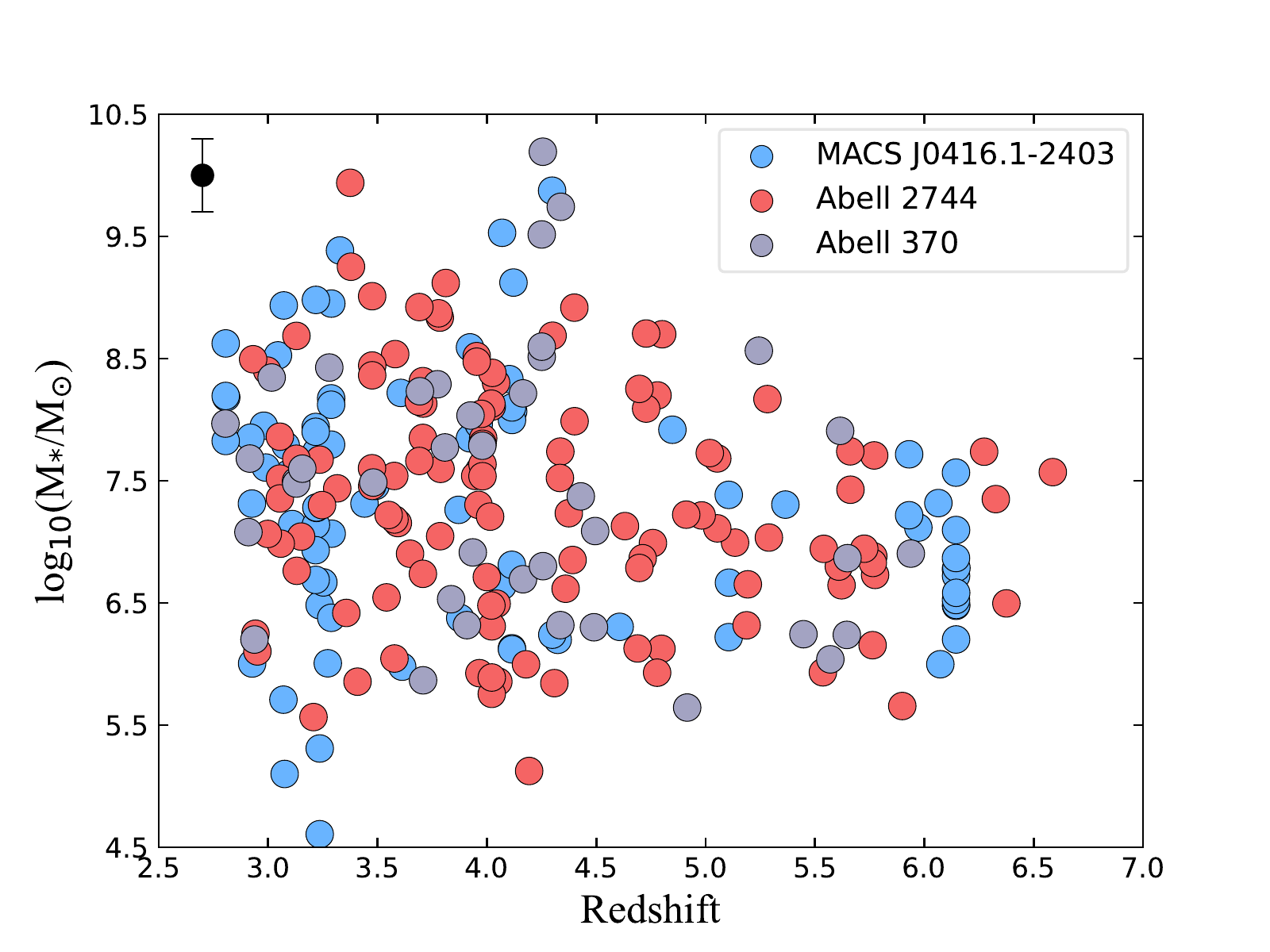}
    \caption{Magnification-corrected stellar masses versus redshift for our HFF Ly$\alpha$ emitters. Stellar masses are corrected for lensing magnification. The black point in the upper left corner indicates the average stellar mass error.}
    \label{Fig_5}
\end{figure}

The magnification-corrected stellar masses we derived for our galaxies are very low, ranging from $10^{5.5}\; \mathrm{M_{\odot}} \lesssim \mathrm{M_{*}} \lesssim  10^{10.5}\; \mathrm{M_{\odot}}$. This range of values is in good agreement with what has been found in the literature in other analysis of (smaller) lensed galaxy samples \citep[e.g.,][]{Karman_2017, Mestric_2022}. We show M$_{*}$ as a function of redshift for our sources in Fig. \ref{Fig_5}. From this figure we see that the complete range of stellar masses is displayed by galaxies at $z<4.5$, while we only see galaxies with M$_{*}\lesssim 10^{8.5} \, \rm  M_{\odot}$ at higher redshifts. This is not surprising, because higher stellar mass galaxies become increasingly rarer at higher redshifts and, therefore, larger area surveys are needed to find them.

These low-mass objects represent a galaxy population almost completely unexplored at these redshifts, with stellar masses even lower than those of the galaxies described in \citet{Santini_2017} ($\mathrm{M_{*, min} \approx 10^{7.5}\;M_{\odot}}$). It might be due to a combination of different causes. For instance, our sources have spectroscopic redshifts, and then our SED fitting solutions are well constrained on that parameter compared to \citet{Santini_2017}'s results which are based on photometric redshifts. We also find that our LAEs have a stellar mass consistent with \citet{Karman_2017}. However, at the same time, we identify LAEs with unprecedented low stellar masses (M$_{*} < 10^{6}$ M$_{\odot}$). This result may be due to a combination of the depth of the MUSE data \citep[e.g., the MUSE Deep Lensed Field centred in the north-east part of M0416,][]{Vanzella_2020} and gravitational lensing effects.

These low stellar masses correspond to those of satellite galaxies found in the local Universe \citep[e.g., ][]{Smith_2019, Wang_2021}, and some are even smaller, introducing the possibility that we are looking at stellar aggregates, in the process of forming proto-galaxies \citep[e.g., ][]{Bromm_2011} or proto globular clusters \citep[][]{Vanzella_2016, Vanzella_2017, Vanzella_2017b, Vanzella_2019, Vanzella_2020}. This highlights the advantage of gravitational lensing for investigating intrinsically faint sources, which could otherwise not be observed with current observational facilities in blank fields.  Investigating the presence of stellar groups is beyond the scope of this paper and, therefore, in the following analysis we will consider that all our lensed objects are galaxies, independently of their derived stellar masses.

\subsection{SFR from UV continuum emission for the HFF Ly$\alpha$ emitters}

We derived the SFRs for our  Ly$\alpha$ emitters independently of their SED fitting by considering their rest-frame UV luminosities (L$_{\nu}$). To do that, we calculate L$_{\nu}$ at a reference wavelength $\lambda_{rest}$ = 2000~{\AA} from the photometry of every galaxy at the filter with closest effective wavelength to $\lambda_{obs} = \lambda_{rest}  \times (1+z)$, where $z$ is the redshift of that galaxy. If the observed wavelength falls in between two passbands, we use the mean flux of those two filters as a proxy for f(2000{\AA}). We correct the UV fluxes for dust extinction following the \citet{Calzetti_2000} reddening law in order to recover the intrinsic UV fluxes. To do that, we adopt E(B$-$V) values from the SED fitting analysis. Then we convert them into a monochromatic luminosity (L$_{\nu}$).

Finally we convert L$_{\nu}$ into an SFR using the prescription given by \cite{Kennicut_1998}:
\begin{equation}
     \mathrm{SFR (M_{\odot}\;yr^{-1})} = 1.4\times 10^{-28} \; \mathrm{L_{\nu}} \mathrm{(erg} \; \mathrm{s ^{-1} Hz^{-1}).}
     \label{K98}
\end{equation}
The Kennicutt's conversion formula (Eq. \ref{K98}) has a scatter of 0.3~dex. Therefore, we propagate that error into our uncertainty on the SFR and take into account that our SFRs have to be corrected by magnification effects. Furthermore, Kennicutt's formula is based on a Salpeter IMF \citep[][]{Salpeter_1955}, while, in this work, we adopt a Chabrier one. We convert our SFRs from a Salpeter IMF to a Chabrier one by multiplication with a factor 0.63 \citep[][]{Madau_2014}. The finally obtained star formation rates, determined from the observed UV continuum fluxes, range from 0.001 to 158.49 M$_{\odot}$\;yr$^{-1}$, with a mean of 2.06 M$_{\odot}$\;yr$^{-1}$.

\section{The SFR - M$_{*}$ and sSFR - M$_{*}$ planes}\label{Section_6}

\subsection{The location of the HFF Ly$\alpha$ emitters}

\begin{figure*}[t!]
    \centering
    \includegraphics[width = 0.49 \textwidth, height = 0.30 \textheight]{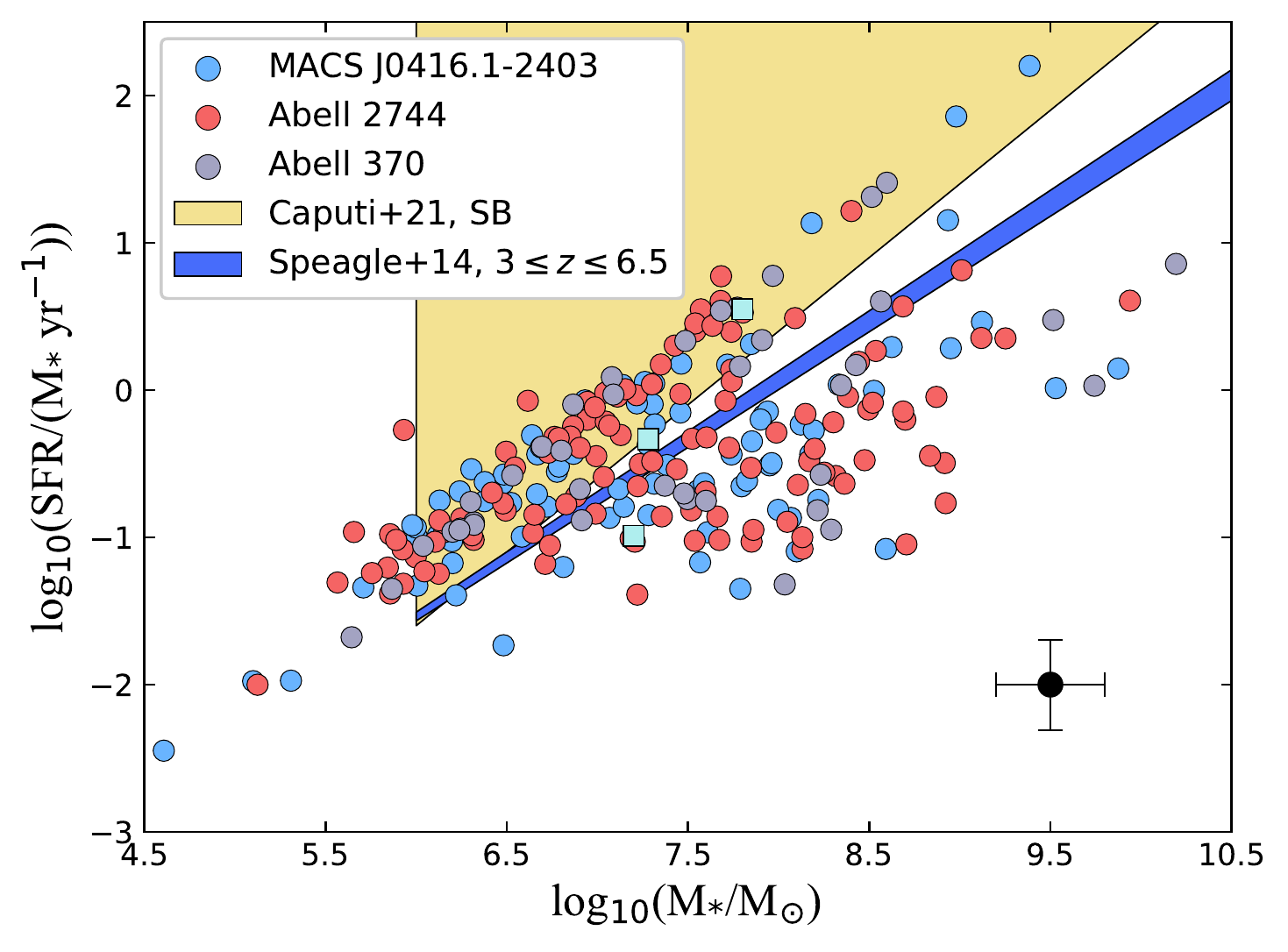}
    \includegraphics[width = 0.49 \textwidth, height = 0.30 \textheight]{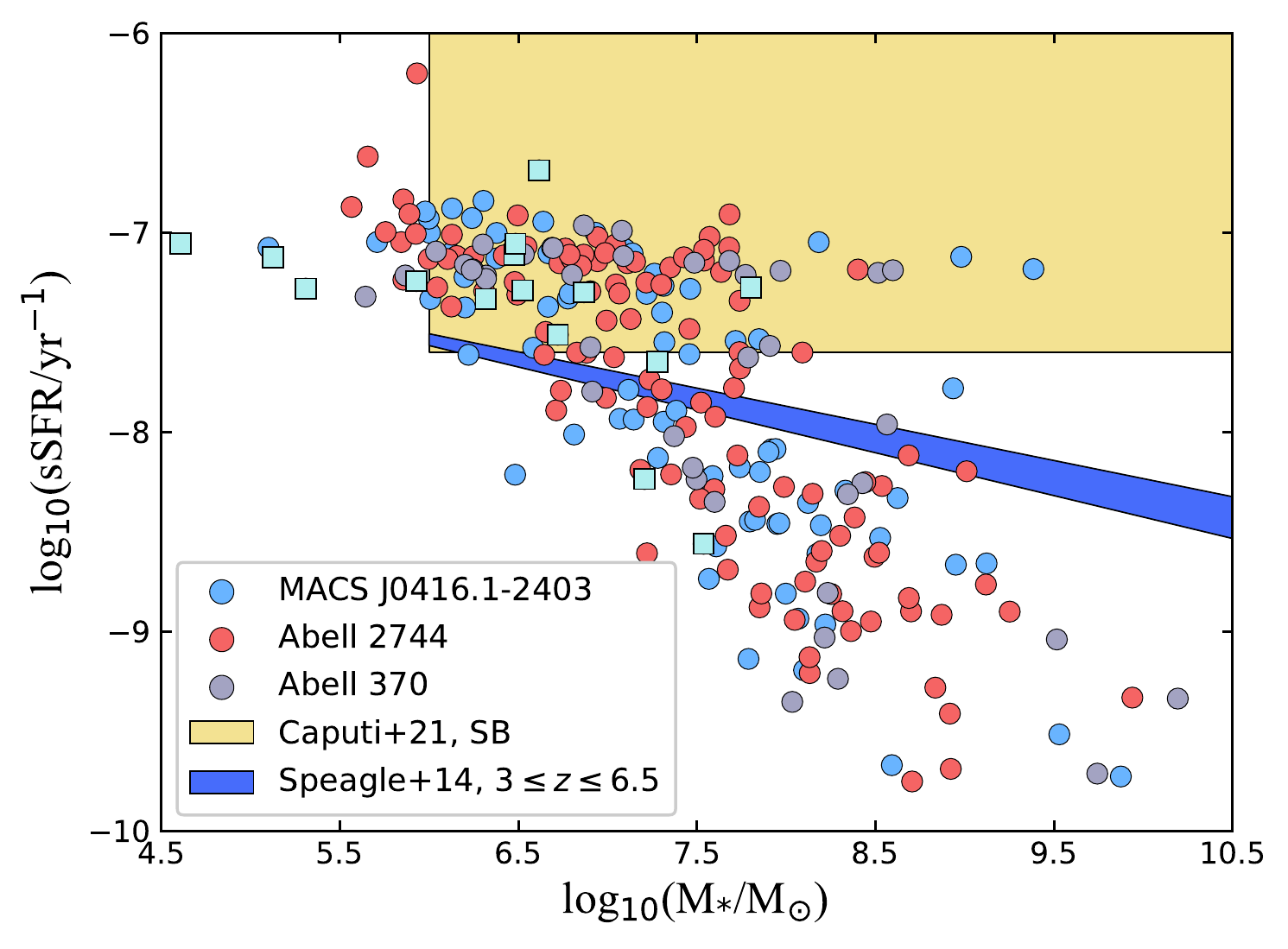}
    \caption{The SFR$-$M$_{*}$ plane (\textit{left}) and the sSFR$-$M$_{*}$ plane (\textit{right}) for our entire sample (M0416 + A2744 + A370) of galaxies at z = 2.8 - 6.5. In both panels,  we compare our data points with the lower envelope of starburst galaxies adopted in \citet{Caputi_2021}. We also show the evolution of Speagle’s MS prescription \citep[][]{Speagle_2014} as a function of redshift ($z = 3 - 6.5$). The error bars indicate average uncertainties. The pale blue squares indicate upper limits.}
    \label{Fig_6}
\end{figure*}

We considered our independent determinations of stellar masses and SFRs to locate our galaxies on the SFR and sSFR versus M$_{*}$  planes (Fig. \ref{Fig_6}). Multiple studies in the literature have shown that most galaxies, on this plane, appear on the so-called {\em star-formation Main Sequence} \citep[e.g., ][]{Brinchmann_2004, Elbaz_2007, Speagle_2014, Whitaker_2014, Salmon_2015}, while a minority lie on a starburst cloud \citep[e.g., ][]{Rodighiero_2011, Caputi_2017, Bisigello_2018}. These previous works only studied galaxies down to $\approx 10^{8.5} \, \rm M_\odot$, while our galaxies probe three decades more down in stellar mass, covering a previously unexplored region in the parameter space.

The location of our lensed Ly$\alpha$ emitters on the SFR$-$M$_{*}$ plane shows that more than half ($\approx 52\%$) of these galaxies are located in the starburst cloud. This statement must be taken with caution for two reasons. First,  until now the starburst cloud has only been determined down to $\approx 10^9 \, \rm M_\odot$ \citep{Caputi_2017} and Fig.~\ref{Fig_6} only shows an extrapolation of the starburst lower envelope \citep{Caputi_2021} toward lower stellar masses. Note, however, that this starburst lower envelope corresponds to stellar-mass doubling times of only  $\approx 4 \times 10^7 \, \rm yr$, which are the typical timescales for local starburst episodes \citep{Knapen_2009}, and therefore classifying all those objects above that envelope as starbursts likely makes sense at any stellar mass.

Secondly, the reason why we do not see galaxies around the extrapolation of the MS is probably because of selection effects. Although these galaxies could exist, they would not be among the Ly$\alpha$ emitters seen by MUSE in lensing fields. Besides, the mere extrapolation of the known MS would cross the starburst envelope at some point at low stellar masses, suggesting that the dichotomy seen at higher stellar masses might not directly apply at lower stellar masses. 

We also investigated whether the magnification corrections could be responsible for the very low stellar masses that we derive for some of the Ly$\alpha$ emitters. Fortunately this is not the case:  if we exclude from our sample those sources with $\mu$ $>$ 10, which constitute $\approx 14\%$ of the total,  we still end up with very low-mass objects (down to $\approx$ 10$^{5.5-6}$ M$_{\odot}$).

\subsection{The analysis of the SFR$-$M$_{*}$ plane over five decades in stellar mass from $z=2.8$ to $z=6.5$}

\begin{deluxetable*}{ccccccc}
\tablenum{3}
\tablecaption{Main Sequence and Starburst best-fit parameters\label{Table_3}.}
\tablewidth{\textwidth}
\tablehead{
\colhead{Redshift} & \colhead{$\alpha$ (MS)} & \colhead{$\beta$ (MS)} & \colhead{$\alpha$ (SB)} & \colhead{$\beta$ (SB)}
}
\startdata
2.8 $\leq$ z $<$ 4 & 0.62 $\pm$ 0.01 & $-$5.18 $\pm$ 0.10  & 1.00 $\pm$ 0.01  & $-$7.29 $\pm$ 0.04 \\
4 $\leq$ z $<$ 5 & 0.60 $\pm$ 0.03 & $-$4.93 $\pm$ 0.31 & 0.99 $\pm$ 0.01  & $-$7.03 $\pm$ 0.10\\
5 $\leq$ z $\leq$ 6.5 & 0.59 $\pm$ 0.05 & $-$4.79 $\pm$ 0.55 & 1.00 $\pm$ 0.02 & $-$7.22 $\pm$ 0.12\\
\enddata
\tablecomments{The errors on $\alpha$ and $\beta$ have been estimated adopting the bootstrap resampling method. }
\end{deluxetable*}
In order to do a more complete study of the SFR$-$M$_{*}$ plane and investigate the evolution of galaxy location with redshift on that plane, we combined our HFF Ly$\alpha$ emitter sample with the COSMOS/SMUVS galaxy sample \citep{Deshmukh_2018, van_Mierlo_2022}. The latter catalogue allows us to incorporate almost 23,000 galaxies at redshifts $z=2.8-6.5$, which mainly populate the plane at stellar masses $\gtrsim 10^9 \, \rm M_\odot$. These galaxies are only the SMUVS star-forming galaxies with a UV-derived SFR, i.e., passive galaxies from \citet{Deshmukh_2018}, as well as star-forming galaxies without the necessary photometric information to compute the UV-based SFRs, have been excluded from our analysis. To do that, we applied the same methodology we used for the HFF Ly$\alpha$ emitters.
\begin{figure*}[t!]
    \centering
    \includegraphics[width = \textwidth]{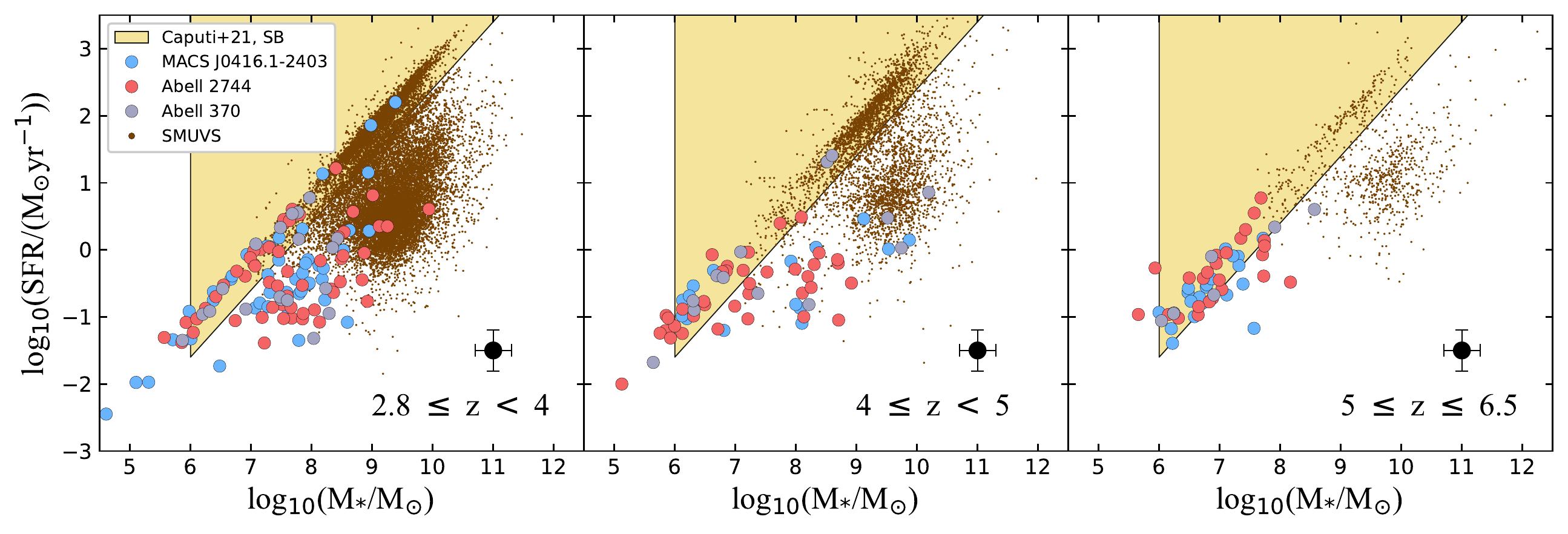}
    \caption{The $\mathrm{SFR}-\mathrm{M_{*}}$ plane, populated with all sources (HFF + SMUVS) considered in this work, divided according to redshift interval as indicated. The error bars indicate the average uncertainty estimates. The yellow region indicates for reference the lower envelope of starburst galaxies adopted in \citet{Caputi_2021}.
    }
    \label{Fig_7}
\end{figure*}
To analyse the redshift evolution, we split the SFR$-$M$_{*}$ plane into three different redshift bins (Fig. \ref{Fig_7}):
\begin{itemize}
    \let\labelitemi\labelitemii
    \item 2.8 $\leq$ z $<$ 4, with 17813 object in total;
    \item 4 $\leq$ z $<$ 5, with 4173 sources in total;
    \item 5 $\leq$ z $\leq$ 6.5,  with 866 galaxies in total.
\end{itemize}
We also analysed the sSFR distribution taking into account both HFF sources and SMUVS ones (Fig. \ref{Fig_8}).
\begin{figure}[t!]
    \centering
    \includegraphics[width = 0.49 \textwidth, height = 0.30 \textheight]{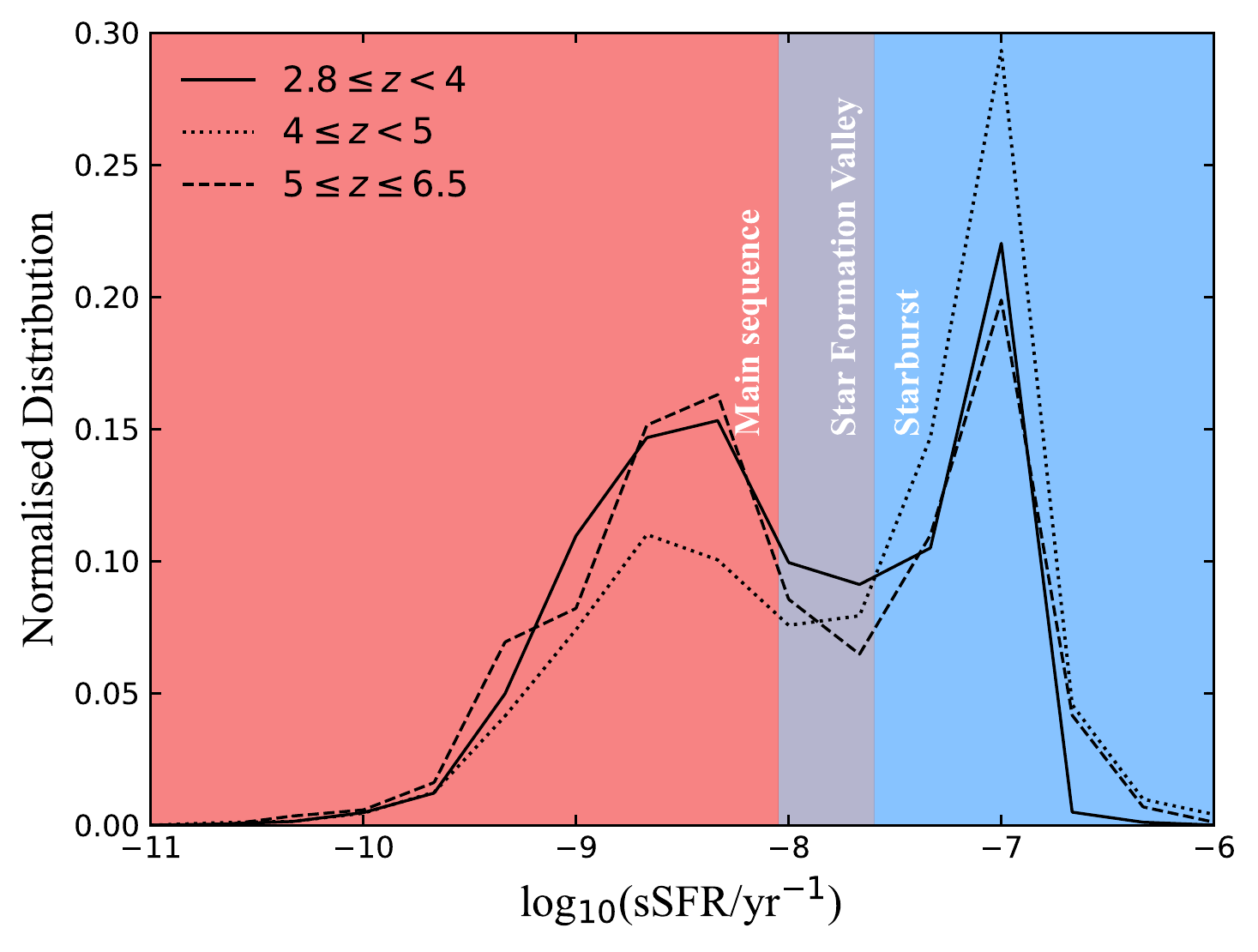}
    \caption{The sSFR distribution of the entire sample (HFF + SMUVS) in each redshift bin. The entire plane is colour coded following the regions derived by \citet{Caputi_2017}: the star-formation MS for sSFR $>$ 10$^{-8.05}$ yr$^{-1}$, the Starburst cloud for sSFR $>$ 10$^{-7.60}$ yr$^{-1}$, and the Star Formation Valley for 10$^{-8.05}$ yr$^{-1}$ $\leq$ sSFR $\leq$ 10$^{-7.60}$ yr$^{-1}$.}
    \label{Fig_8}
\end{figure}

First of all, we recover the same starburst/MS bimodality for star-forming galaxies found by \citet{Caputi_2017}.  This result is not trivial, as  \citet{Caputi_2017} analysed only H$\alpha$ emitters at $z\approx 4-5$ for which the SFRs were based on the inferred intensities of the H$\alpha$ line. Our results indicate that the starburst/MS bimodality applies to all star-forming galaxies and is independent of the method to infer the SFRs.  The region of the SFR$-$M$_{*}$ plane between these two sequences is sparsely populated, and it corresponds to a {\em star-formation (SF) valley}.  From \citet{Caputi_2017}, we divide our entire sample (HFF+SMUVS) into these following three populations:
\begin{itemize}
    \let\labelitemi\labelitemii
    \item \textbf{SB galaxies}: sSFR $>$ 10$^{-7.60}$ yr$^{-1}$;
    \item \textbf{MS galaxies}: sSFR $<$ 10$^{-8.05}$ yr$^{-1}$;
    \item \textbf{SF valley galaxies}: 10$^{-8.05}$ yr$^{-1}$ $\leq$ sSFR $\leq$ 10$^{-7.60}$ yr$^{-1}$.
\end{itemize}
We perform a linear regression for the SB and MS galaxies separately in each of our considered redshift bins. We fit the MS and starburst sequence adopting the following linear relation:
\begin{equation}
     \mathrm{log_{10}(SFR/M_{\odot} yr^{-1})} = \alpha\mathrm{log_{10}(M_{*}/M_{\odot})} + \beta
     \label{linear_regression}
\end{equation}

We obtained the errors on the slopes and intercepts through the bootstrap resampling method. We iterate the bootstrap resampling 1000 times creating a distribution of $\alpha$ and $\beta$ values, and adopt the standard deviation of these distributions as the $1\sigma$ error on those parameters. In particular, to take into account any possible effect of stellar mass completeness that could affect our sample in the different redshift bins that we adopted, we decided to perform the linear regressions considering different stellar-mass cuts.

For the MS galaxies we do not find any significant differences in the slope and intercept values for the different stellar-mass cuts we investigated (i.e., at 10$^{8}$ M$_{\odot}$, 10$^{8.5}$ M$_{\odot}$, 10$^{9}$ M$_{\odot}$, and 10$^{9.5}$ M$_{\odot}$). We applied the same methodology to SB galaxies. However, in this case, we considered more stellar-mass cuts than for MS galaxies (i.e., down to  M$_{*, min}$ $=$ 10$^{6}$ M$_{\odot}$). Again, we do not find any significant variations even if we go down to very low stellar-mass cuts. To perform the linear regression in the three redshift bins we analysed in this paper, we decided to consider the following stellar-mass cuts for MS galaxies:
\begin{itemize}
    \let\labelitemi\labelitemii
    \item $\mathrm{log_{10}(M_{*}/M_{\odot}) \geq 8}$ at $2.8 \leq z < 4$;
    \item $\mathrm{log_{10}(M_{*}/M_{\odot}) \geq 8.5}$ at $4 \leq z < 5$;
    \item $\mathrm{log_{10}(M_{*}/M_{\odot}) \geq 9}$ at $5 \leq z \leq 6.5$.
\end{itemize}
Regarding SB galaxies, we decided to exploit the unparalleled opportunity offered by the gravitational lensing effect to reach low stellar-mass objects, opting for a unique stellar-mass cut at any redshift bins we analysed in this work (i.e., $\mathrm{log_{10}(M_{*}/M_{\odot})} \geq 6$). The results of the linear regressions we performed are listed in Table \ref{Table_3}.

\begin{figure}[t!]
    \centering
    \includegraphics[width = 0.49 \textwidth, height = 0.30 \textheight]{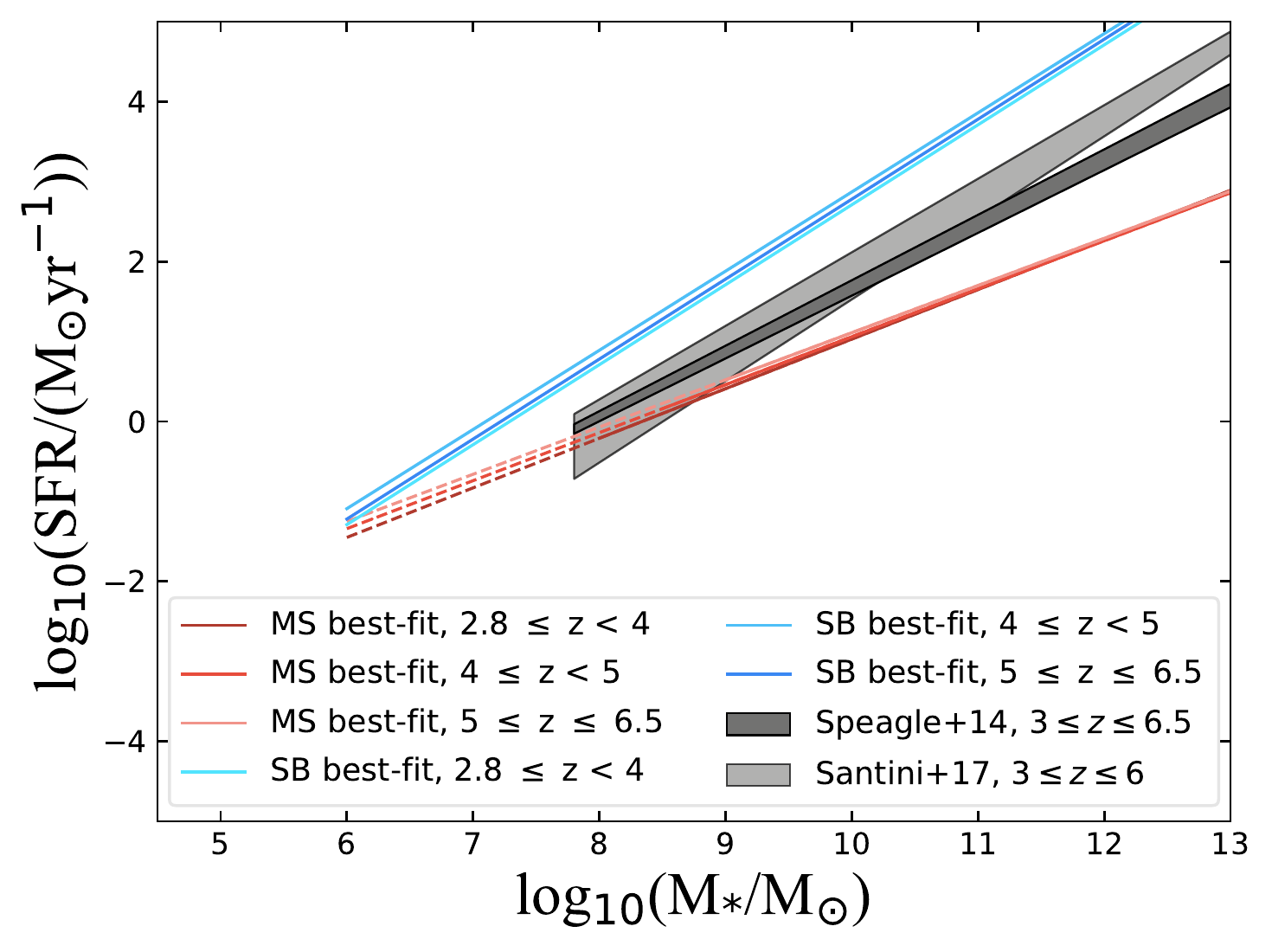}
    \caption{Comparison between our best-fit galaxy Main Sequences with main-sequence determinations from the recent literature: \citet{Speagle_2014} and \cite{Santini_2017}. For them, we show the evolution of the MS in the same redshift interval we analyse in this work. For the MS linear regressions, we adopted three different mass-cuts for the three redshift bin we studied: 10$^{8}$ M$_{\odot}$ for $2.8 \leq z < 4$, 10$^{8.5}$ M$_{\odot}$ for $4 \leq z < 5$, and 10$^{9}$ M$_{\odot}$ for 5 $\leq$ z $\leq$ 6.5. We show their extrapolation down to 10$^{6}$ M$_{\odot}$ adopting dashed lines.  We also show our starburst cloud linear regressions for reference, highlighting there is no redshift evolution for them within the error bars. }
    \label{Fig_9}
\end{figure}

In Fig. \ref{Fig_9} we show our separate main sequence and starburst cloud fittings at $z = 2.8-4$, $z = 4-5$, and $z =5-6.5$. We find that there is no or very marginal evolution in $\alpha$ for the MS galaxies. Furthermore, we do not see a redshift evolution in the SB sequence. Finally, we recover a large value of $\beta$ in the MS at $z = 5 - 6.5$ if compared with the lowest redshift bins. However, we do not observe an evolution within the error bars between $z=4-5$ and $z=5-6.5$. 

Furthermore, we analysed if the presence of the significant peak of sources at $z\approx 6$ in MACS J0416.1-2403 (Fig. \ref{Fig_5}) could affect our results. They are 11 sources in total. We analysed their $\mathrm{M_{*}}$, SFRs and sSFRs. All of them have $\mathrm{M_{*} < 10^{8}\;M_{\odot}}$. One of them falls in the SF valley galaxies (10$^{-8.05}$ yr$^{-1}$ $\leq$ sSFR $\leq$ 10$^{-7.60}$ yr$^{-1}$). The other ones have $\mathrm{sSFR > 10^{-7.60}\; yr^{-1}}$. For this reason, we classify all of them as starburst galaxies. Since they are starbursts and at $z\approx 6$, we repeated the linear regression for starburst galaxies in the third redshift bin ($5 \leq z \leq 6.5$) without taking them into account. We do not find any significant difference compared to what we show in Table \ref{Table_3}. Therefore, we conclude that including them does not affect our results.

To put our results in context, we compared them with the most recent literature about the Main Sequence of star-forming galaxies. In Fig.~\ref{Fig_9}  we also show the evolution of the MS determinations from \citet{Speagle_2014, Santini_2017} in our same redshift intervals. Note that, as pointed out by \citet{Caputi_2017}, the comparison must be done with care. Indeed, it could lead to misunderstandings since many authors did not apply the net separation between MS and SB galaxies we adopted in this work \citep[e.g.,][]{Rodighiero_2011}. As we can see in Fig. \ref{Fig_9}, both \citet{Speagle_2014} and \citet{Santini_2017} MS prescriptions fall between the MS and SB curves we obtained in this work. This result might be due to the fact that both \citet{Speagle_2014} and \citet{Santini_2017} do not apply any separation between MS and SB galaxies as we did in this work. Therefore, as we stated above, the comparison of results must be done with care.

\subsection{The role of starburst galaxies at z $>$ 3}

\begin{deluxetable*}{cccccccccc}
\tablenum{4}
\tablecaption{Number of sources (HFF + SMUVS) for each population, i.e., MS, SB and SF Valley (SFV), in the SFR$-$M$_{*}$ plane\label{Table_4}.}
\tablewidth{\textwidth}
\tablehead{
\colhead{Redshift} \vspace{-0.2cm} & \colhead{M$_{*}$} & \colhead{MS} & \colhead{SB} & \colhead{SFV} & \colhead{MS fraction} & \colhead{SB fraction} & \colhead{SFV fraction} \\
\colhead{bin} & \colhead{($\mathrm{log_{10}(M_{*}/M_{\odot})}$)} & \colhead{} & \colhead{} & \colhead{} & \colhead{(\%)} & \colhead{(\%)} & \colhead{(\%)}
}
\startdata
2.8 $\leq$ z $<$ 4 & 4.5 $\leq \mathrm{M_{*}}<$ 5.5 & 0 & 3 & 0 & --- & --- & --- \\
 & 5.5 $\leq \mathrm{M_{*}} <$ 6.5 & 1 & 16  & 0 & --- & ---  & ---  \\
 & 6.5 $\leq \mathrm{M_{*}} <$ 7.5 & 6 & 84  & 12 & --- & --- &  --- \\
 & 7.5 $\leq \mathrm{M_{*}} <$ 8.5 & 252 & 723  & 425 & 18.00  & 51.64 & 30.36  \\
 & 8.5 $\leq \mathrm{M_{*}} <$ 9.5 & 5735 & 3764  & 1348 & 52.87  & 34.70 & 12.43  \\
 & 9.5 $\leq \mathrm{M_{*}} <$ 10.5 & 3897 & 894  & 453 & 74.31  & 17.05  &  8.64\\
 & 10.5 $\leq \mathrm{M_{*}} <$ 11.5 & 190 & 4  & 6 & 95.00 & 2.00  & 3.00 \\
 & $\mathrm{M_{*}}\geq$ 11.5 & 0 & 0  & 0 & 0.00  & 0.00  &  0.00\\
 \hline
  & Total & 10081 & 5488 & 2244 & 56.59 & 30.81 &  12.60 \\
\hline
4 $\leq$ z $<$ 5 & 4.5 $\leq \mathrm{M_{*}} <$ 5.5 & 0 & 1  & 0 & --- & --- & --- \\
 & 5.5 $\leq \mathrm{M_{*}}<$ 6.5 & 0 & 19  & 0 & ---  & --- & ---\\
 & 6.5 $\leq \mathrm{M_{*}}<$ 7.5 & 2 & 33  & 12 & --- & --- & --- \\
 & 7.5 $\leq \mathrm{M_{*}}<$ 8.5 & 33 & 241  & 64 & --- & --- & --- \\
 & 8.5 $\leq \mathrm{M_{*}}<$ 9.5 & 554 & 1148  & 252 &  28.35 & 58.75  & 12.90  \\
 & 9.5 $\leq \mathrm{M_{*}}<$ 10.5 & 1033 & 512  & 158 & 60.66  & 30.06 & 9.28 \\
 & 10.5 $\leq \mathrm{M_{*}}<$ 11.5 & 85 & 10  & 8 & 82.52 & 9.71 & 7.77 \\
 & $\mathrm{M_{*}}\geq$ 11.5 & 2 & 0  & 1 & 66.67  & 0.00  & 33.33  \\
 \hline
  & Total & 1715 & 1964 & 494 & 41.10 & 47.06 & 11.84  \\
\hline
5 $\leq$ z $\leq$ 6.5 & 4.5 $\leq \mathrm{M_{*}}<$ 5.5 & 0 & 0  & 0 & --- & --- & --- \\
 & 5.5 $\leq \mathrm{M_{*}} < $ 6.5 & 0 & 12  & 1 & --- & --- & ---  \\
 & 6.5 $\leq \mathrm{M_{*}} < $ 7.5 & 0 & 22  & 6 & ---  & --- & --- \\
 & 7.5 $\leq \mathrm{M_{*}} < $ 8.5 & 4 & 73  & 8 & --- & ---  & --- \\
 & 8.5 $\leq \mathrm{M_{*}} < $ 9.5 & 95 & 133  & 47 & 34.55 & 48.36 & 17.09 \\
 & 9.5 $\leq \mathrm{M_{*}} < $ 10.5 & 339 & 49  & 20 & 83.09 & 12.01 & 4.90  \\
 & 10.5 $\leq \mathrm{M_{*}} < $ 11.5 & 50 & 1  & 1 & 96.16  & 1.92  & 1.92 \\
 & $\mathrm{M_{*}}\geq$ 11.5 & 6 & 0  & 0 & 100.00  & 0.00  & 0.00  \\
 \hline
  & Total & 493 & 290 & 83 & 56.93  & 33.49  & 9.58  \\
  \hline
  \hline
  \colhead{Redshift} \vspace{-0.2cm} & \colhead{} & \colhead{MS} & \colhead{SB} & \colhead{SFV} & \colhead{MS fraction} & \colhead{SB fraction} & \colhead{SFV fraction} \\
\colhead{bin} & \colhead{} & \colhead{} & \colhead{} & \colhead{} & \colhead{(\%)} & \colhead{(\%)} & \colhead{(\%)}\\
\hline
  2.8 $\leq$ z $\leq$ 6.5 & & 12289 & 7742 & 2821 & 53.77 & 33.87 & 12.36  \\
\enddata
\tablecomments{We report both the number and fraction of MS, SB, and SF valley galaxies in each redshift bin and stellar mass bin (1 dex). We also report the total number and total fraction of each population in each redshift bin we adopted in this work. Finally, we show the total amount of MS, SB, and SF valley galaxies, and their fractions, from $z = 2.8$ to $z = 6.5$. Blank rows refer to the stellar-mass bins that are significantly incomplete.}
\end{deluxetable*}

Over the past decades, the possible existence of different regions on the SFR$-$M$_{*}$ plane, as a consequence of different modes of star formation \citep[e.g., ][]{Daddi_2010, Genzel_2010, Elbaz_2011}, has been studied extensively from the local Universe \citep[][]{Renzini_2015} to z $\approx$ 3 \citep[][]{Santini_2009, Rodighiero_2011, Bisigello_2018,  Daddi_2013, Ilbert_2015}. \citet{Daddi_2010} suggested that star formation occurs through two different regimes: a long-lasting mode for disks and a more rapid mode for starbursts. There is a broad consensus regarding that MS galaxies grow up on a long time scale as a consequence of a smooth gas accretion from the Intergalactic medium \citep[e.g.,][]{Almeida_2014, Renzini_2015, Pearson_2018}. On the contrary, the nature of SB galaxies is still under debate, and many theories have been proposed to describe them: violent disk instability \citep[][]{Inoue_2016, Romeo_2016, Tadaki_2018}, merger events \citep[e.g.,][]{Sanders_1988, Elbaz_2003, Lamastra_2013, Calabro_2019}, and even a sort of primeval galaxies with a high amount of total gas \citep[][]{Scoville_2014, Genzel_2015, Scoville_2017}. In particular, the role of starburst galaxies in the cosmic history of star formation is not completely understood as well \citep[e.g., ][]{Sargent_2012}. For instance, \citet{Rodighiero_2011} pointed out that, between $z = 1.5$ and $z = 2.5$, SB galaxies represent only 2\% of their sample which account for only 10\% of the SFR density (SFRD) in that redshift interval. However, their conclusion is based on massive galaxies only (M$_{*}$ $>$ 10$^{10}$ M$_{\odot}$). In this work, instead, we cover a much wider range in stellar mass and show evidence of a prominent starburst sequence along with the Main Sequence on the SFR$-$M$_{*}$ plane in three different redshift bins, from z $\approx$ 2.8 to z $\approx$ 6.5 (Fig. \ref{Fig_7}), consistent with the findings by \citet{Caputi_2017} at $z\approx 4-5$. 

\begin{figure*}[t!]
    \centering
    \includegraphics[width =\textwidth]{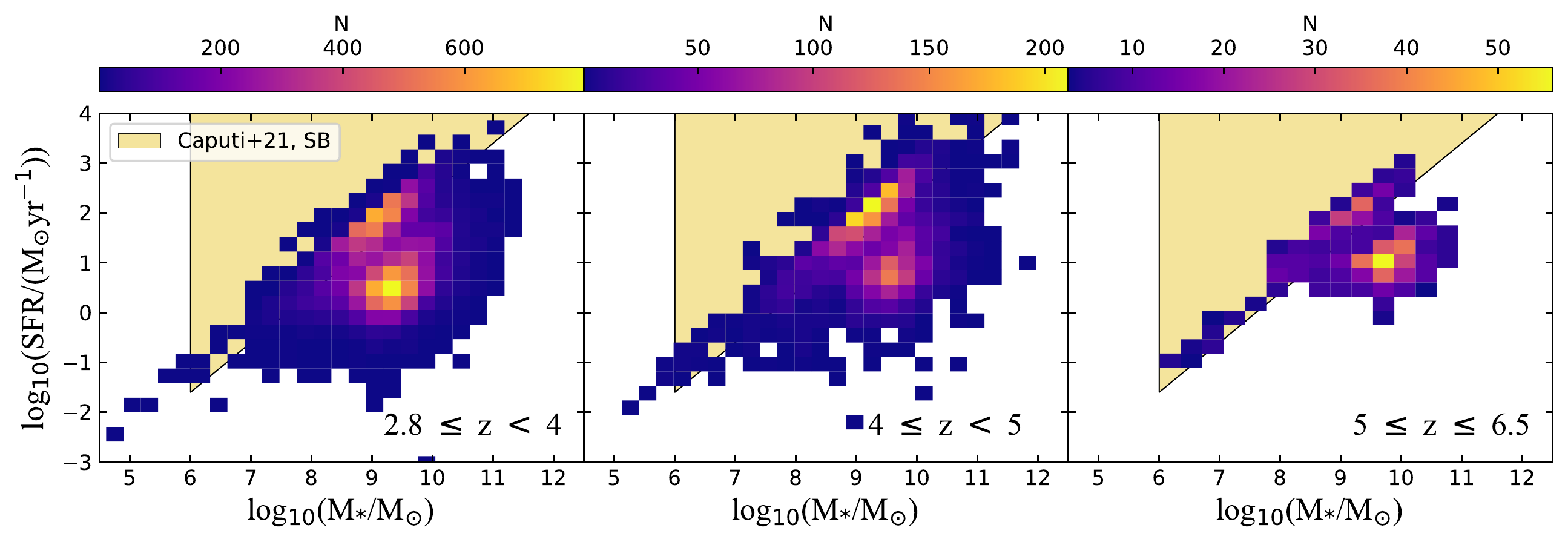}
    \caption{The 2D distributions showing the galaxy distributions (HFF + SMUVS) in the $\mathrm{SFR-M_{*}}$ plane in the three redshift bins adopted in this work, as indicated. The same two prominent features are present in each redshift interval: the starburst cloud, and the Main Sequence. For reference the lower envelope of starburst galaxies adopted in \citet[][]{Caputi_2021} is also shown.}
    \label{Fig_10}
\end{figure*}

In Fig. \ref{Fig_10}, we show the 2D distributions of the SFR$-$M$_{*}$ plane as a function of redshift, considering the three redshift bins that we adopted in this work. Our aim with this plot is to quantify the fraction of galaxies in the MS and the SB cloud at each redshift. These 2D distributions clearly show that the starburst/MS bimodality discussed above is present at all our analysed redshifts.

Moreover, we analysed the fractions of MS, SB, and SF valley galaxies, as well as their contribution to the total SFR budget, in each of the redshift bin we adopted in this work. A detailed compilation of the number of galaxies for each population (MS, SB, and SF valley), at different stellar masses and redshifts,  is shown in Table \ref{Table_4}.

\begin{figure*}[t!]
    \centering
    \includegraphics{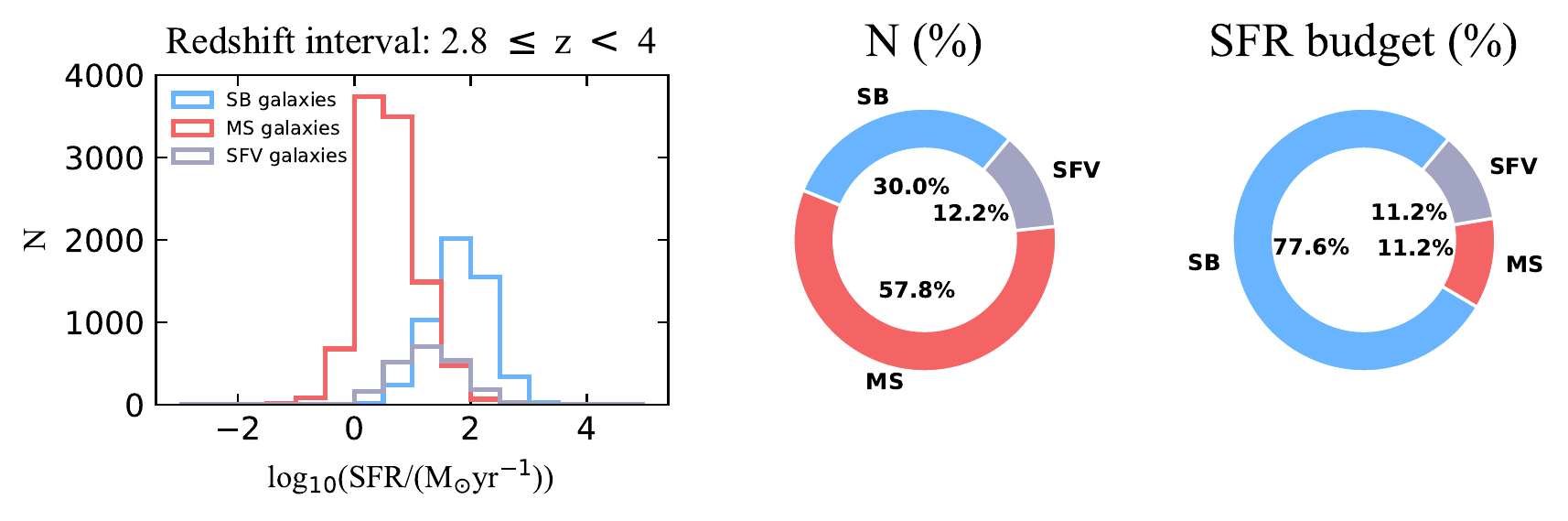}
    \caption{\textbf{Left panel}: SFR distribution of MS, SB, and SF valley galaxies at $z = 2.8 - 4$. \textbf{Central panel}: A pie chart showing the fraction of MS, SB, and SF valley galaxies at $z = 2.8 - 4$. \textbf{Right panel}: A pie chart showing the percentages contributed to the overall SFR budget by these three galaxy populations. In these plots, we adopted the same stellar-mass cut for MS, SB, and SF valley galaxies at $\mathrm{M_{*} \geq 10^{8}\;M_{\odot}}$.}
    \label{Fig_11}
\end{figure*}

In Fig. \ref{Fig_11}, we show the fraction of MS, SB, and SF valley galaxies at $z = 2.8 - 4$ considering only galaxies with $\mathrm{log_{10}(M_{*}/M_{\odot}) \geq 8}$, where all the three populations are stellar-mass complete at $>70-80\%$ level. We find a high fraction of starburst galaxies ($30\%$), much larger than what has been found in the previous literature from the local Universe to  high redshifts \citep[e.g., ][]{Rodighiero_2011, Bergvall_2016, Schreiber_2015, Caputi_2017}. The large fraction of starburst galaxies we find here is dominated by the low stellar-mass galaxies. Remarkably, these starbursts make for $\approx 80\%$ of the total SFR budget at these redshifts, indicating the importance of the starburst star-formation mode.

\begin{figure*}[p]
    \centering
    \includegraphics{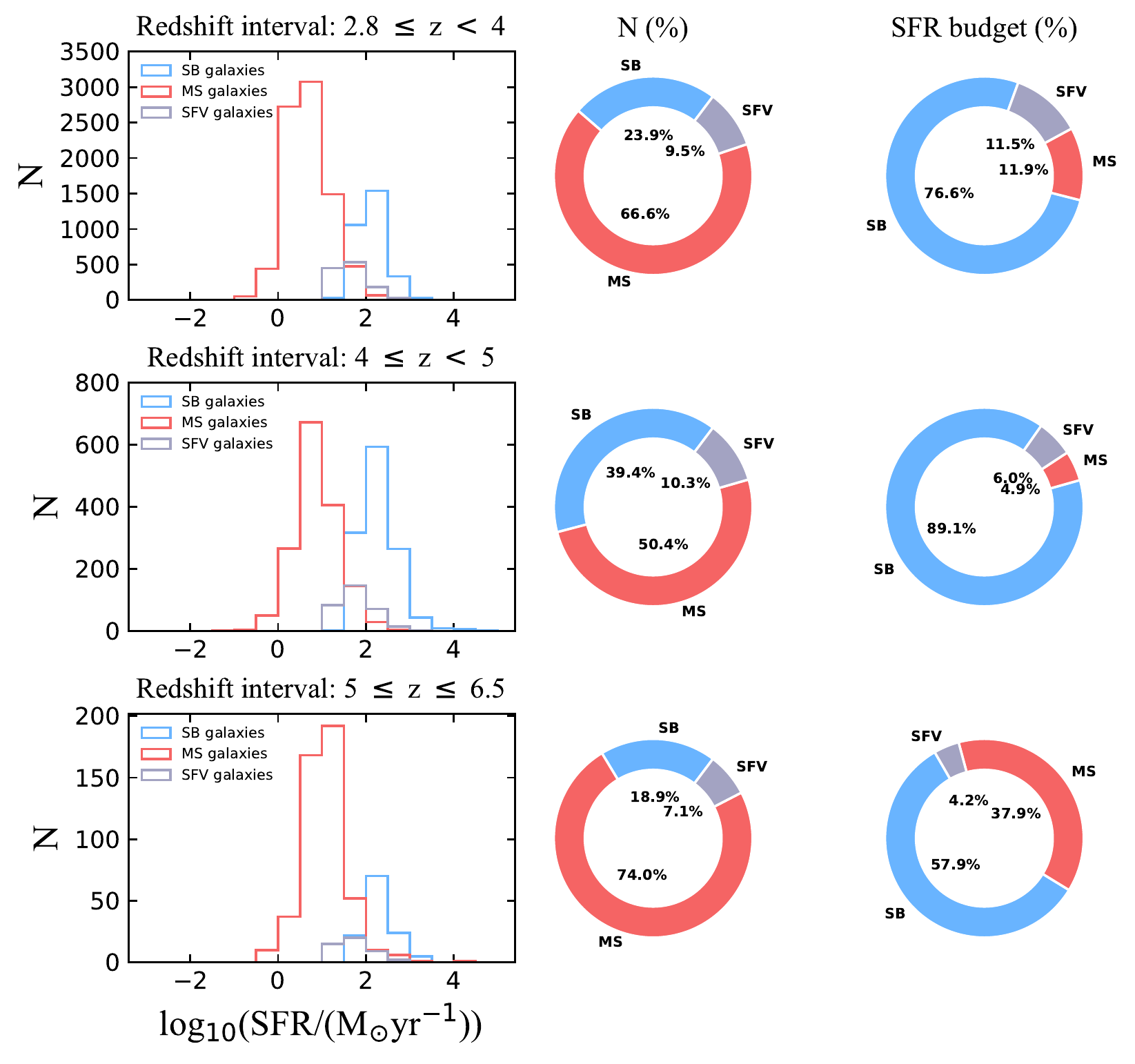}
    \caption{\textbf{Left panels}: SFR distributions of MS, SB, and SF valley galaxies with $\mathrm{M_{*} \geq 10^{9}\;M_{\odot}}$, at different redshift bins. \textbf{Central panels}: Pie charts showing the corresponding fractions of MS, SB, and SF valley galaxies. \textbf{Right panels}: Pie charts showing the percentages that each population contributes to the overall SFR budget. }
    \label{Fig_12}
\end{figure*}

We also investigated the evolution of the starburst fraction as a function of redshift. To do that, we adopted the same stellar-mass cut ($\mathrm{log_{10}(M_{*}/M_{\odot}) \geq 9}$) for all redshift bins, in order to ensure high stellar-mass completeness even at the highest redshifts. In Fig. \ref{Fig_12}, we show the fractions of MS, SB, and SF valley galaxies with $\mathrm{log_{10}(M_{*}/M_{\odot}) \geq 9}$ in the three redshift bins, i.e., $z = 2.8 - 4$, $z = 4 - 5$, and $z= 5 - 6.5$. 

On the one hand, we find a significantly higher fraction of starburst galaxies at $z = 4-5$ than at other redshifts, suggesting the existence of a preferential epoch for the starburst phenomenon. On the other hand, we notice a dramatic drop of this fraction at $z > 5$ among the $\mathrm{log_{10}(M_{*}/M_{\odot}) \geq 9}$ star-forming galaxies. This highlights again that the starburst phenomenon preferentially occurs at low stellar-masses ($\mathrm{M_{*} \lesssim 10^{9}\;M_{\odot}}$).

Furthermore, we evaluated if including sources selected as passive galaxies in the SMUVS sample (see Section 2.2) could really affect our conclusions. To investigate this point,  we considered all SMUVS galaxies (without separating them into star-forming and passive), and all HFF galaxies, and simply made a cut in sSFR:  $\mathrm{sSFR > 10^{-9.8}\;yr^{-1}}$, as suggested in \citet[][]{Bisigello_2018}. Then, we estimated the fraction of MS, SB and SF valley galaxies at $z = 2.8 - 4$ considering only galaxies with $\mathrm{log_{10}(M_{*}/M_{\odot}) \geq 8}$, as we did in Fig \ref{Fig_11}. In this case, we find that MS galaxies account for 58.7\% instead of 57.8\%. Starburst galaxies represent 29.2\% of total amount of sources instead of 30\%. Finally, SF valley galaxies account for 12.1\% instead of 12.2\%. This result makes clear that excluding passive galaxies does not affect our results at all.

We estimated the uncertainties on the fractions of MS, SB, and SF valley galaxies that we show in Fig \ref{Fig_12}, via Monte Carlo Markov Chain (MCMC) simulations. We created 1000 mock catalogues from our initial sample (HFF + SMUVS). To do that, we perturb $\mathrm{M_{*}}$ and $\mathrm{SFR}$ within their error bars for each run of MCMC simulations. Moreover, for the SMUVS sources, we decide to perturb redshifts as well because SMUVS galaxies have photometric redshifts in contrast to HFF sources. To do that, we analyse the redshift probabilty distribution (PDZ) of the SMUVS galaxies. In particular, we focus on those sources that show a secondary peak solution ($\mathrm{z_{sec}}$) with a non-zero probability ($\mathrm{P(z_{sec}) = 1-P(z_{best})}$). We construct our MCMC simulations in order to randomly choose between $\mathrm{z_{best}}$ and $\mathrm{z_{sec}}$ (when a secondary peak solution exists, i.e., $\mathrm{P(z_{sec})>0}$) adopting their probabilities as a weight. After we create our mock catalogues, we group mock galaxies according to the three redshift bins we analysed in this work. Once we have our final mock catalogues, we split the mock galaxies in MS, SB and SF valley objects, applying the same sSFR criteria explained in Section 6.2. Then, for each mock catalogue, we estimate the fraction of MS, SB, and SF valley galaxies in all the redshift bins. As a result, for each redshift interval, we construct a distribution of fractions for MS, SB, and SF valley galaxies. Finally, we define the 1$\sigma$ error on those fractions as the half-distance between the 16th and 84th percentile of each distribution. At $z = 2.8 - 4$ and $z = 4 - 5$, we recover, for each population, an error of $\approx 1\%$. These errors rise up to $\approx 2\%$ at $z = 5 - 6.5$.

Another result shown in Fig. \ref{Fig_12} is that, regardless the percentage of galaxy types we found in each redshift bin, the majority of the total SFR, even if we apply a stellar-mass cut at $\mathrm{M_{*} \geq 10^{9}}$,  is always produced by the starbursts. That is particularly true for the second redshift bin since, as showed above, the majority of galaxies at $z = 4 - 5$ are located in the starburst cloud, regardless the stellar-mass cut we decided to adopt. This result is extremely important because it demonstrates that starburst galaxies have had an important role in cosmic star formation history, particularly over the first few billion years of cosmic time.

\subsection{Implications for the cosmic star formation rate density}

We use our derived SFRs to obtain an estimate of the cosmic SFR density at z $\approx$ 2.8 $-$ 6.5. To do that, in each redshift interval, we sum SFRs up and then divide them by the comoving volume encompassed by that redshift bin\footnote{We estimate the corresponding comoving volume for the SMUVS survey (i.e., 0.66 sq. deg.). We obtained the comoving volume for the entire sky with the Cosmo calculator at \url{http://www.astro.ucla.edu/~wright/CosmoCalc.html.}}.

In particular, we consider the SMUVS sources only to estimate the SFR density. We decided not to consider the HFF galaxies since they are lensed objects and the computation of the effective volume probed in these fields introduces an additional source of uncertainties from the lens models. Nonetheless, we estimated that the contribution of the HFF galaxies to the total SFR (for HFF + SMUVS) in each redshift bin is negligible ($\lesssim 1\%$).

\begin{figure*}[t!]
    \centering
    \includegraphics[width =  \textwidth]{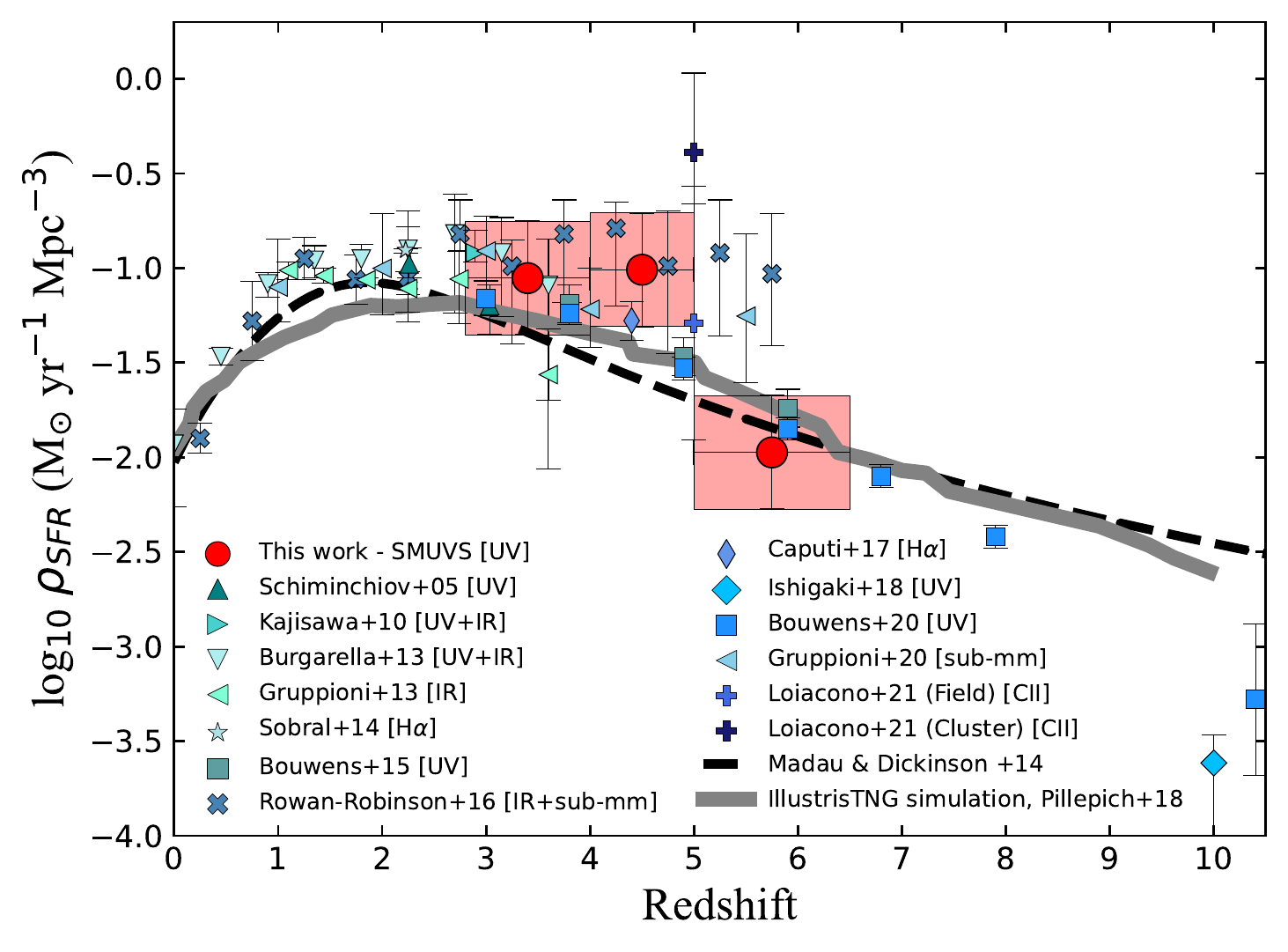}
    \caption{Cosmic star formation rate density versus redshift. The large red circles indicate our estimates in the three redshift bins we adopted in this work. To derive those quantities, we estimated the total SFR in each redshift bin, and then we divided that by the corresponding comoving volume. We did not correct our SFRD estimates for incompleteness, which makes them lower limits. All these estimates have been done without taking into account the AGN fraction in the SMUVS catalogue at $z \approx$ 2.8$-$6.5. Other symbols refer to the recent SFRD determinations from the literature, based on different SFR tracers
    \citep[][]{Schiminovich_2005, Kajisawa_2010, Burgarella_2013, Gruppioni_2013, Sobral_14, Bouwens_2015, Rowan_Robinson_2016, Caputi_2017,  Ishigaki_2018, Bouwens_2020, Gruppioni_2020, Loiacono_2021}. The different curves correspond to theoretical predictions. Dashed line: \citet{Madau_2014}. Solid line: \citet{Pillepich_2018}.  All the SFRD values in this figure correspond to a \citet{Chabrier_2003} IMF over stellar masses (0.1$-$100)M$_{\odot}$.
    }
    \label{Fig_13}
\end{figure*}

In Fig. \ref{Fig_13}, we show the redshift evolution of the SFRD, the so-called {\sl Lilly-Madau diagram} \citep[][]{Lilly_1996, Madau_1996}. In this plot, we included our derivations of SFRD in the three redshift bin we adopted, as well as a compilation of recent results in the literature which are based on different SFR tracers estimated exploiting different galaxy surveys and individual galaxy SFR:
\begin{itemize}
\let\labelitemi\labelitemii
    \item UV: \citet{Schiminovich_2005}, \citet{Bouwens_2015}, \citet{Ishigaki_2018}, and \citet{Bouwens_2020};
    \item Emission lines: \citet{Sobral_14}, \citet{Caputi_2017} and \citet{Loiacono_2021};
    \item IR and sub-mm: \citet{Gruppioni_2013}, \citet{Rowan_Robinson_2016}, and \citet{Gruppioni_2020};
    \item Combination of UV and IR: \citet{Kajisawa_2010} and \citet{Burgarella_2013}.
\end{itemize}

The SFRD estimates we shows in Fig. \ref{Fig_13} are not corrected for incompleteness, which makes them lower limits.

We find a nearly flat distribution at $z = 2.8 - 5$. We find a good agreement between our results and those from the recent literature we show, albeit those results are based on different SFR tracers. Moreover, we find a significant decline of the SFRD at $z = 5 - 6.5$, which is probably due to the fact we miss obscured objects or we are not able to properly recover the correct dust-extinction factors, but in agreement with \cite{Bouwens_2015, Bouwens_2020} found in their work.

Many additional studies have provided SFRD estimations at high redshift in the previous literature (Fig. \ref{Fig_14}). The majority of them, like ours, are based on UV fluxes. However, it is crucial to remember that an estimate based solely on UV fluxes might be heavily influenced by dust-extinction corrections, which are higher than those for other SFR tracers. As a result, their estimations are subject to significant uncertainty, particularly at high redshifts \citep[e.g.,][]{Castellano_2014}.

In spite of these plausible uncertainties, it is clear from  Fig. \ref{Fig_13} that our resulting SFRD value at $z=4-5$ is significantly higher than most previous determinations from the literature (as compiled by \citet{Madau_2014}; dashed line) and consistent, within the errors, with far-IR-derived SFRD  \citep[e.g.,][]{Gruppioni_2013, Rowan_Robinson_2016} and line-emission based SFRD \citep[][]{Caputi_2017, Loiacono_2021}. Our inferred SFRD is also significantly above the predictions of the \texttt{IllustrisTNG} simulations \citep[gray solid curve; ][]{Pillepich_2018}. At redshifts up to $\approx 5 - 5.5$, we observe that both the SFRD from the literature and our own derivations have a sharp decline. This is likely the effect of incompleteness -- many dust-obscured and/or low-mass galaxies could be missing at such high redshifts.

\subsection{The comparison between observations and simulations: IllustrisTNG}

We decided to compare our results with the recent \texttt{\href{https://www.tng-project.org/about/}{IllustrisTNG}} simulations. The \texttt{IllustrisTNG} simulations \citep[][]{Marinacci_2018, Naiman_2018, Nelson_2018, Pillepich_2018, Springel_2018} are cosmological magneto-hydrodynamical simulations with the purpose of reproducing processes considered extremely important in the field of galaxy formation and evolution. These simulations are the result of an improvement of the original Illustris project \citep[e.g., ][]{Nelson_2015} by including new models such as a new black-hole feedback model, magneto-hydrodynamics, a new scheme for galactic winds, and many other features. The initial conditions of these simulations have been initialised at $z = 127$ and the cosmological assumptions are based on \citet{Planck_2016}. 

Over the past three years, a number of increasing outcomes have been published to show the agreement between observations and these simulations \citep[e.g., ][]{Genel_2018}. As a result, \texttt{IllustrisTNG} simulations provide an excellent laboratory in which to compare our findings. In this work, we adopted the TNG50 simulation which is the last simulation of the \texttt{IllustrisTNG} project and corresponds to a cosmological box with a side length of 50~$\rm Mpc/h$  \citep{Nelson_2019}. We focused on two important galaxy properties, namely M$_{*}$ and SFR, since we want to compare the observed SFR $-$ M$_{*}$ plane with that derived from the cosmological models. 

\begin{figure*}[ht!]
    \centering
    \includegraphics[width = \textwidth]{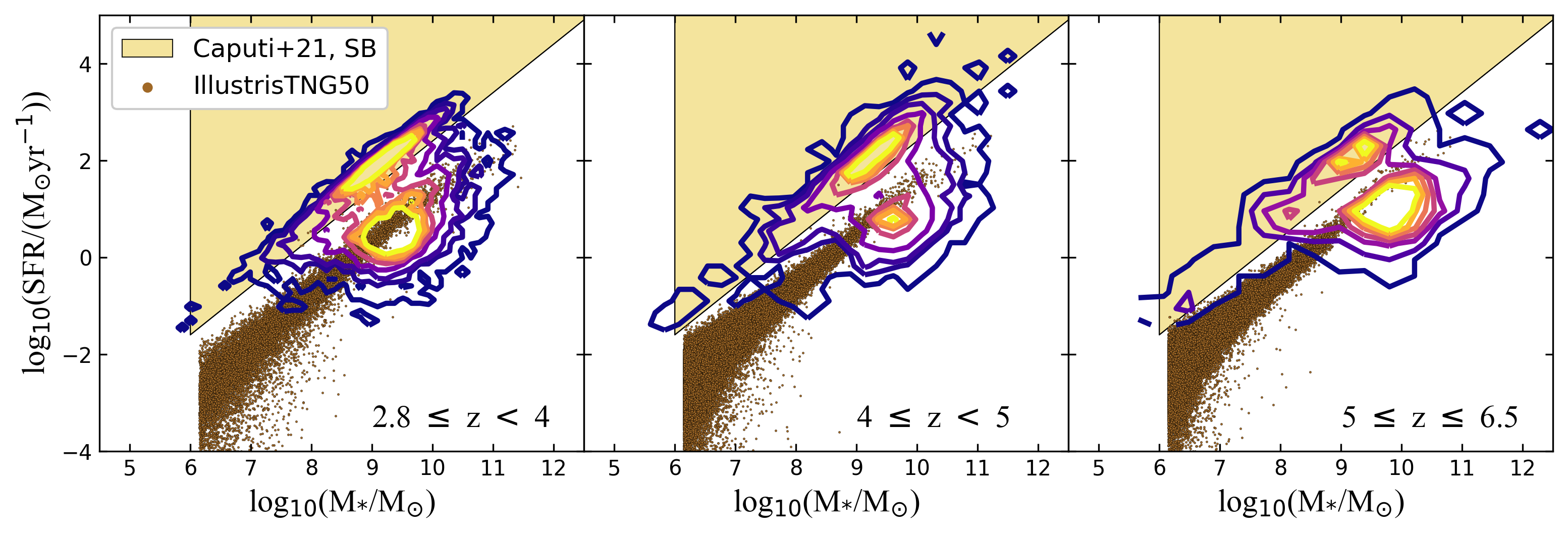}
    \caption{The $\mathrm{SFR}-\mathrm{M_{*}}$ plane. We show the comparison between observations and simulations. The contour plots refer to our entire sample (HFF + SMUVS). The brown points refer to \texttt{IllustrisTNG50} galaxies. We show the $\mathrm{SFR}-\mathrm{M_{*}}$ plane in the three redshift bins we adopted. We also show for reference the lower envelope of starburst galaxies adopted in \citet{Caputi_2021}}
    \label{Fig_14}
\end{figure*}
In Fig. \ref{Fig_14} we show the SFR $-$ M$_{*}$ plane where we compare our sample (HFF + SMUVS) with the \texttt{IllustrisTNG50} one, in the three redshift bins that we analysed throughout this work. We can clearly see in spite of probing a cosmological volume which is significantly larger than the one probed by COSMOS/SMUVS, \texttt{IllustrisTNG50} does not predict the presence of a starburst cloud. Instead, \texttt{IllustrisTNG50} galaxies lie on a unique tight relation crossing the upper part of our MS. 
 
\begin{figure}[t!]
    \centering
    \includegraphics[width = 0.49 \textwidth, height = 0.30 \textheight]{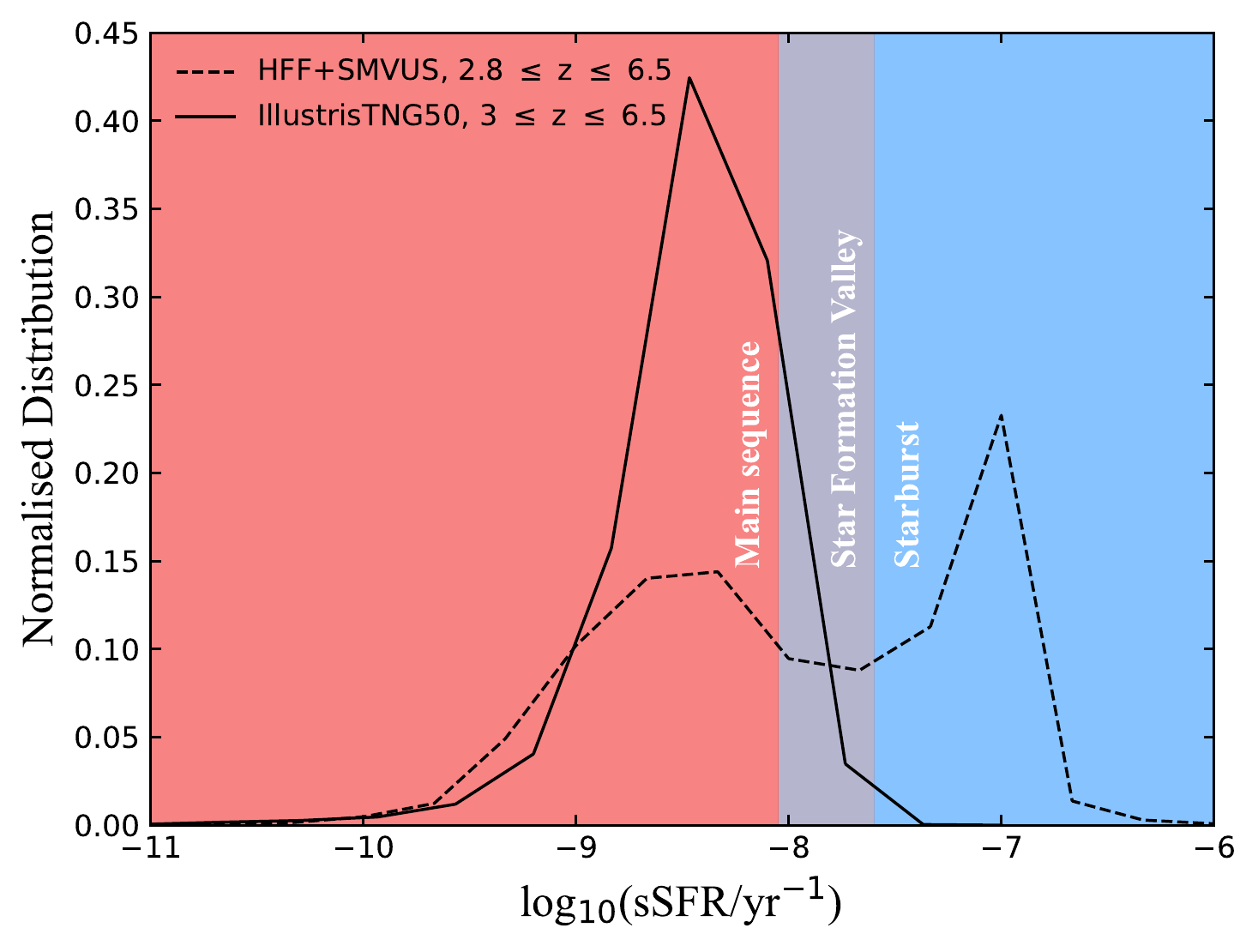}
    \caption{We show the comparison of the sSFR distribution between our entire sample (dashed line) and the IllustrisTNG one (solid line). The entire plane is colour coded following the prescription of \citet{Caputi_2017}: the Main sequence for sSFR $>$ 10$^{-8.05}$ yr$^{-1}$, the Starburst cloud for sSFR $>$ 10$^{-7.60}$ yr$^{-1}$, and the Star Formation Valley for 10$^{-8.05}$ yr$^{-1}$ $\leq$ sSFR $\leq$ 10$^{-7.60}$ yr$^{-1}$.}
    \label{Fig_15}
\end{figure}

This lack of starburst galaxies can more clearly be seen in the sSFR distribution (Fig. \ref{Fig_15}). Indeed, if we look at the comparison between the sSFR distributions of our sources versus the \texttt{IllustrisTNG50} ones, we see that the simulated galaxies do not show any kind of bimodality as we find from the observations. A similar result has been found by \citet{Katsianis_2021}, although their study is focused on the dichotomy between star-forming and passive galaxies \citep[see also][]{Zhao_2020, Caballero_2021}. The lack of starburst galaxies in theoretical galaxy models was also pointed out for previous generations of galaxy formation models \citep[e.g., ][]{Sparre_2015}. This starburst absence may be due to the insufficient resolution of galaxy models, as the physics involved in the starburst phenomenon could occur at very small length scales, which are not properly resolved in  state-of-the-art simulations.

\section{Discussion and conclusion}\label{Section_7}

We have investigated the relation between SFR and M$_{*}$ over five decades in M$_{*}$, with a joint analysis of $\mathrm{\approx 23,000}$ star-forming galaxies from the COSMOS/SMUVS galaxy survey and a sample of 240 lensed LAEs from three HFF lensing galaxy clusters (M0416, A2744, and A370),  all at $2.8 \leq z \leq 6.5$. We have derived and analysed their stellar properties. In particular, we considered the rest UV fluxes to estimate the SFR and sSFR for all of these galaxies.

The LAEs that we analysed here have a stellar-mass ranging from $10^{5.5}\; \mathrm{M_{\odot}} \lesssim \mathrm{M_{*}} \lesssim  10^{10.5}\; \mathrm{M_{\odot}}$, providing an unparalleled chance to study star formation in low stellar-mass objects (Fig. \ref{Fig_6}). We still trace stellar masses down to $\mathrm{M_{*}\approx 10^{5.5-6}\; M_{\odot}}$ even if we only consider galaxies with modest magnification values ($\mu < 10$).

We found that more than a half of our LAEs at $z=3-6.5$ lie on the starburst cloud ($\approx 52\%$). This trend is particularly noticeable at low stellar masses. This is likely the consequence of a selection effect: low stellar-mass galaxies with lower star-formation rates may exist, forming the extrapolation of the star-formation MS, but are undetected with current telescopes. Interestingly, the ages derived from the best-fit SEDs of our low stellar-mass SB are comparable to their stellar-mass doubling times and the starburst phenomenon typical timescales ($\mathrm{\approx 10^{7}\;yr}$, \citealp{Heckman_2006}), suggesting that we are catching these galaxies in their first starburst episode, i.e., in the process of being formed. In the near future, the {\em James Webb Space Telescope}  will allow us to probe whether older low stellar-mass galaxies exist at these high redshifts.

The SMUVS galaxies  mainly populate the SFR$-$M$_{*}$ plane at stellar masses $\gtrsim 10^{9}\;\mathrm{M_{\odot}}$. in this regime, we found a similar starburst/MS bimodality (Fig. \ref{Fig_8}) to that discovered  by  \citet{Caputi_2017}. This bimodality recovery is non-trivial:  \citet{Caputi_2017} only analysed a sample of H$\alpha$ emitters at $z \approx 4 - 5$ and used a different SFR tracer. Our finding is reassuring, as it demonstrates that the presence of starburst/MS bimodality does not depend on which SFR tracer is adopted.

We also investigated the evolution of SFR$-$M$_{*}$ plane as a function of redshift at $z=2.8-6.5$. Following \citet{Caputi_2017}, we split the SFR$-$M$_{*}$ plane into three regions (SB galaxies, MS galaxies, and SF valley galaxies).  We found basically no evolution in the MS slope, in agreement with previous works. There is no evolution in its normalization either, within the error bars, which instead seems at odds with most of the previous literature.  However, as pointed out by \citet{Caputi_2017}, the direct comparison of the MS slope and normalisation with most previous works could be misleading, as they typically do not segregate starburst galaxies in their studies, so their star-formation MS appears artificially elevated.  For the SB sequence, we also found that it does not evolve at all with redshift within the error bars. 

Our results indicate that starbursts constitute more than 20\% of all star-forming galaxies with $\mathrm{M_{*} \gtrsim 10^{9}\;M_{\odot}}$ at $2.8 \leq z \leq 6.5$ and reach a peak of 40\% at $z=4-5$ (Fig. \ref{Fig_12}),  suggesting that this redshift range corresponds to a preferential epoch for the starburst phenomenon \citep[][]{Faisst_2019, Atek_2022, Vanderhoof_2022}. More importantly, although MS galaxies outnumber SB galaxies, we found that, at all redshifts, the majority of the star formation rate budget is produced by the SB (Fig. \ref{Fig_12}). These results differ from what has been found in most of the previous literature since prior studies only looked at galaxies with $\mathrm{M_{*}} \gtrsim 10^{10}\;\mathrm{M_{\odot}}$, without taking into account the contribution of low-mass star-forming galaxies. In this work, we also show the implications of these results on the overall cosmic SFRD, strongly suggesting, as we stated above, that starburst galaxies  played a crucial role in the first few billion years of cosmic time (Fig. \ref{Fig_13}), in agreement with the recent conclusion by \citet{Asada_2021}.

We have also compared our results with the predictions of the \texttt{IllustrisTNG50} galaxy simulations. In this work, we show that, at all the redshift intervals that we investigated, simulations cannot reproduce the starburst cloud that we identify from observations (Fig. \ref{Fig_14}). Furthermore, galaxy simulations do not predict any starburst/MS bimodality. This result is quite clear if we consider the sSFR distribution for observed and simulated galaxies (Fig. \ref{Fig_15}), suggesting a plausible lack of resolution in galaxy models, as the physics involved in the starburst phenomenon could occur at very small length scales, which the current hydrodynamical simulations cannot probe. Further observational constraints 
of the starburst stellar and gas contents should help improve this aspect of galaxy models.

\begin{acknowledgments}
We thank an anonymous referee for a careful reading and useful comments on this manuscript.

Based in part on observations carried out with the Spitzer Space Telescope, which is operated by the Jet Propulsion Laboratory, California Institute of Technology under a contract with NASA. Also based on data products from observations conducted with ESO Telescopes at the Paranal Observatory under ESO program ID 179.A-2005 and on data products produced by TERAPIX and the Cambridge Astronomy Survey Unit on behalf of the UltraVISTA consortium. Also based on observations carried out by NASA/ESA Hubble Space Telescope, obtained and archived at the Space Telescope Science Institute; and the Subaru Telescope, which is operated by the National Astronomical Observatory of Japan. This research has made use of the NASA/IPAC Infrared Science Archive, which is operated by the Jet Propulsion Laboratory, California Institute of Technology, under contract with NASA.

KIC, SvM, GBC and EI acknowledge funding from the European Research Council through the award of the Consolidator Grant ID 681627-BUILDUP. GBC acknowledge the Max Planck Society for financial support through the Max Planck Research Group for S. H. Suyu and the academic support from the German Centre for Cosmological Lensing.

\end{acknowledgments}

\vspace{5mm}
\facilities{{\sl HST}, VLT/MUSE, {\sl Spitzer}, {\sl VISTA}, {\sl Subaru}}.

\software{\texttt{Astropy} \citep{astropy_2018}, 
          \texttt{LePHARE} \citep{LePhare_2011},
          \texttt{Photutils} \citep{Photutils},
          \texttt{Source Extractor} \citep{SExtractor},
          \texttt{TOPCAT} \citep{Topcat}.
          }

\bibliography{References}{}
\bibliographystyle{aasjournal}

\end{document}